\documentclass[useAMS,usenatbib]{mn2e}

\usepackage{epsfig}
\usepackage{textcomp}
\usepackage{amssymb,amsmath}
\usepackage{longtable}
\usepackage{lscape}
\usepackage{hyperref}

\newcommand{\xmm} {{\it XMM-Newton}}
\newcommand{\chandra} {{\it Chandra}}

\newcommand{\cmsq} {cm$^{-2}$}
\newcommand{\nh} {$N_{\rm{H}}$}
\newcommand{\lx} {$L_{\rm{X}}$}

\newcommand{\oiii}{{\rm{[O\,\sc{iii}]}}}

\newcommand{\degree}{{$^\circ$}}
\newcommand{\ergs}{\mbox{\thinspace erg\thinspace s$^{-1}$}}
\newcommand{\ergcms}{\mbox{\thinspace erg\thinspace cm$^{-2}$\thinspace s$^{-1}$}}

\newcommand{\lamedd} {$\lambda_{\rm Edd}$}
\newcommand{\fscatt} {$f_{\rm scatt}$}

\newcommand{\nctagn}{   157}
\newcommand{\ngctagn}{   100}
\newcommand{\ngctagna}{    30}
\newcommand{\ngctagnb}{    42}
\newcommand{\ngctagnc}{    28}
\newcommand{\nhctagn}{    22}

\newcommand{\ncctagn}{    18}

\newcommand{\npctagn}{   345}
\newcommand{\nctagnhiz}{     5}
\newcommand{\ctfrac}{   5.7}
\newcommand{\ctfraca}{   9.5}
\newcommand{\ctfracb}{   7.0}
\newcommand{\ctfracc}{   4.1}
\newcommand{\cteffa}{  69.7}
\newcommand{\cteffb}{  98.2}
\newcommand{\cteffc}{  39.5}

\newcommand{\nhabstor}{   925}

\newcommand{\nsrc}{  3184}
\newcommand{\nsrca}{   561}
\newcommand{\nsrcb}{   923}
\newcommand{\nsrcc}{  1700}
\newcommand{\nred}{  3109}
\newcommand{\nreda}{   538}
\newcommand{\nredb}{   923}
\newcommand{\nredc}{  1648}
\newcommand{\nzspeca}{   349}
\newcommand{\nzspecb}{   463}
\newcommand{\nzspecc}{   850}
\newcommand{\zcomp}{  97.6}
\newcommand{\zcompa}{  95.9}
\newcommand{\zcompb}{ 100.0}
\newcommand{\zcompc}{  96.9}
\newcommand{\nstarsa}{     8}
\newcommand{\nstarsb}{    14}
\newcommand{\nstarsc}{    61}

\newcommand{\meangamml}{  1.71$_{-  0.04}^{+  0.03}$}
\newcommand{\sdgamml}{  0.26$_{-  0.04}^{+  0.04}$}
\newcommand{\nhabs}{   760}
\newcommand{\nunobs}{  1913}
\newcommand{\nunobsa}{   290}
\newcommand{\nunobsb}{   490}
\newcommand{\nunobsc}{  1133}
\newcommand{\meanlx}{ 43.18}
\newcommand{\meanlxa}{ 42.58}
\newcommand{\meanlxb}{ 43.01}
\newcommand{\meanlxc}{ 43.49}
\newcommand{\meanfscatt}{  1.79}
\newcommand{\gamtora}{  1.59$\pm$  0.14}
\newcommand{\gamtorb}{  1.72$\pm$  0.15}
\newcommand{\gamtorc}{  1.70$\pm$  0.08}

\newcommand{\nhicntsa}{   156}
\newcommand{\nlocntsa}{   393}
\newcommand{\nbigcdifa}{    65}
\newcommand{\nhifsca}{    19}
\newcommand{\nhrabsa}{    15}

\newcommand{\nhicntsb}{    65}
\newcommand{\nlocntsb}{   872}
\newcommand{\nbigcdifb}{   155}
\newcommand{\nhifscb}{    24}
\newcommand{\ncthifxb}{     5}
\newcommand{\nhrabsb}{    13}

\newcommand{\nhicntsc}{    25}
\newcommand{\nlocntsc}{  1736}
\newcommand{\nbigcdifc}{   207}
\newcommand{\nhifscc}{    40}

\newcommand{\nhrabsc}{     4}

\title[]{Compton thick active galactic nuclei in Chandra surveys}

\author[M. Brightman et al]{Murray Brightman$^{1}\thanks{E-mail: mbright@mpe.mpg.de}$, Kirpal Nandra$^{1}$, Mara Salvato$^{1}$, Li-Ting Hsu$^{1}$, Cyprian Rangel$^{2}$\\
$^{1}$Max-Planck-Institut f\"{u}r extraterrestrische Physik, Giessenbachstrasse 1, D-85748, Garching bei M\"{u}nchen, Germany\\
$^{2}$Astrophysics Group, Imperial College London, Blackett Laboratory, Prince Consort Road, London SW7 2AZ, UK\\}

\begin{document}

\maketitle

\label{firstpage}

\begin{abstract}
We present the results from an X-ray spectral analysis of active galactic nuclei (AGN) in the \chandra\ Deep Field-South, AEGIS-XD and \chandra-COSMOS surveys, focussing on the identification and characterisation of the most heavily obscured, Compton thick (CT, \nh$>10^{24}$ \cmsq) sources. Our sample is comprised of \nsrc\ X-ray selected extragalactic sources, which has a high rate of redshift completeness (\zcomp\%), and includes improved photometric redshifts over previous studies. We use spectral models designed for heavily obscured AGN which self consistently include all major spectral signatures of heavy absorption. We identify CT sources not selected through our spectral fitting method using X-ray colours, validate our spectral fitting method through simulations, and take considerations for the constraints on \nh\ given the low count nature of many of our sources. After these considerations we identify a total of \ngctagn\ CT AGN with best-fit \nh$>10^{24}$ \cmsq\ and \nh\ constrained to be above $10^{23.5}$ \cmsq\ at 90\% confidence. These sources cover an intrinsic 2-10 keV X-ray luminosity range of $10^{42} - 3\times10^{45}$ \ergs\  and a redshift range of z=0.1-4. This sample will enable characterisation of these heavily obscured AGN across cosmic time and to ascertain their cosmological significance. These survey fields are sites of extensive multi-wavelength coverage, including near-infrared CANDELS data and far-infrared {\it Herschel} data, enabling forthcoming investigations into the host properties of CT AGN. Furthermore, by using the torus models to test different covering factor scenarios, and by investigating the inclusion of the soft scattered emission, we find evidence that the covering factor of the obscuring material decreases with \lx\ for all redshifts, consistent with the receding torus model, and that this factor increases with redshift, consistent with an increase in the obscured fraction towards higher redshifts. The strong relationship between the parameters of obscuration and \lx\ points towards an origin intrinsic to the AGN, however the increase of the covering factor with redshift may point towards contributions to the obscuration by the host galaxy. We make \nh, $\Gamma$ (with uncertainties), observed X-ray fluxes and intrinsic 2-10 keV luminosities for all sources analysed in this work publicly available in an online catalogue.

\end{abstract}

\section{Introduction}

Active galactic nuclei (AGN) are widely accepted to be powered by the accretion of material onto a supermassive black hole (SMBH) in the centres of galaxies via the release of gravitational potential energy. This accretion power is one of the dominant energy generation mechanisms in the universe. Not only is this accretion process interesting in the realms of extreme physics, but the growth of a galaxy's nuclear SMBH is believed to be linked to the growth of the galaxy itself, despite the vastly different scales involved \cite[e.g.][]{silk98}. 

Much of this power, however, is obscured from sight due to intervening material. This has been shown directly from observations by measurement of the line of sight absorption in X-ray spectra \citep{awaki91,risaliti99}, and has also been inferred, due to a large population of obscured AGN being required to fit the cosmic X-ray background \citep[CXB, ][]{comastri95,ueda03,gilli07,ueda14}. The obscuration is posited to be due to a torus-like structure surrounding the nucleus \citep[e.g.][]{antonucci85}. This torus is the basis of orientation dependent unified schemes, which explain the differences between type 1 and type 2 AGN \citep[e.g.][]{antonucci93,urry95}. 

X-ray observations of AGN are particularly useful in tracing accretion power due to their unique capability to penetrate all but the largest of columns of gas and dust. X-rays are generated in AGN via the inverse-Compton scattering of optical/UV photons, which are produced by thermal emission in the accretion disc, by hot electrons forming a hot corona in the vicinity of the disc \citep[e.g.][]{sunyaev80}.

Due to the large fraction of AGN which are obscured, characterising and understanding the obscuration is important for determining the accretion history of the universe \citep{fabian99} and accurate determination of the AGN X-ray luminosity function \citep[e.g.][]{aird10,ueda14}. Furthermore, an obscured growth phase of AGN is predicted by some models of coeval galaxy/SMBH formation, suggesting that obscured AGN are a key stage in the process \citep[e.g.][]{hopkins06}. So far studies of the obscured AGN population have not revealed the decisive link between obscured accretion and galaxy properties, such as morphology, star formation rate or stellar mass \citep[e.g.][]{schawinski12,rosario12,rovilos12,merloni14}. However, it is possible that the missing link lies with the extremely obscured, Compton thick population, which may well have been unaccounted for in these studies so far \citep{kocevski14}. In the local universe, most CT AGN reside in spiral galaxies \citep{goulding12}, with only a few exceptions, such as the early type galaxy, ESO 565ÐG019 \citep{gandhi13}, where the authors suggest the CT AGN in this galaxy may have been triggered by a minor merger.

\subsection{Compton thick AGN}

Compton thick AGN are characterised by extreme flux suppression, where the line of sight column density, \nh, exceeds $1.5\times10^{24}$ \cmsq. Due to this, these sources are not only difficult to detect, but at times difficult to distinguish from  less absorbed AGN due to low signal to noise spectra. In the local universe only tens of secure CT AGN are known \citep[e.g.][]{risaliti99,bassani99,matt00,guainazzi05,goulding12}. The fraction of AGN shown to be Compton thick at low redshift has generally been found to be $\sim20$\% \citep{burlon11,brightman11b} from hard X-ray and mid-infrared (MIR) selected samples. Studies at higher redshift are more difficult but there is evidence that this increases to $\sim40$ \% above $z=1$ \citep{brightman12b}. While a significant fraction of CT AGN is required to fit the CXB \citep{comastri95,gilli07}, this fraction is largely unconstrained by the CXB, with statistically acceptable fractions ranging from 5-50\%  \citep{akylas12}.

Many works concentrating on the characterisation of Compton thick AGN have used indirect methods to identify them. These often use high ratios of mid-infrared to optical (MIR/O) or mid-infrared to X-ray (MIR/X) fluxes to infer heavy obscuration, assuming that the MIR flux originates from reprocessed dust emission, and that the low optical or X-ray flux with respect to this is due to heavy obscuration \citep[e.g.][]{daddi07,fiore08}. However, contamination of the MIR by star formation processes can be problematic in these studies \citep{georgakakis10,rangel13}. X-ray hardness ratios are commonly used to infer a Compton thick column. These often also miss CT AGN, however, due to soft X-ray emission emanating from scattering or thermal emission from larger radii, leading to softer hardness ratios than expected for simple obscuration \citep{guainazzi05,brightman12}.

Arguably the most robust method of identifying CT AGN is through X-ray spectroscopy. \nh\ can usually be directly measured in the radiation transmitted through the medium by the position of the photoelectric absorption turn-over. When the medium becomes optically thick to Compton scattering (\nh$>1.5\times10^{24}$ \cmsq), the spectrum becomes dominated by X-ray reflection, which is characterised by a flat observed spectrum ($\Gamma\sim1$) accompanied by an intense iron K$\alpha$ line \citep{matt96_2}. The flat spectral shape is produced by a combination of reprocessing of the primary emission by photoelectric absorption and Compton scattering from a medium which is optically thick to Compton scattering, where the line of sight to the source is obscured. This means that CT AGN can be identified via X-ray spectroscopy even if all direct emission is suppressed. 

When spectral fitting, it is commonplace to fit each of these components individually. While this is suitable for high signal to noise spectra afforded by long pointed observations of nearby AGN, the low signal to noise spectra from observations of the distant Universe using deep surveys does not make this practical. In recent years, new spectral models have been produced that self-consistently account for all of the major signatures of heavy obscuration, with the assumption of a torus geometry \citep{murphy09,ikeda09,brightman11}. These are better suited to low signal to noise spectra due to the fewer number of free parameters. This was demonstrated by \cite{brightman12b} using the \cite{brightman11} model in fitting CDFS spectra, identifying 41 CT AGN. In addition to being well suited to low signal to noise spectra, these models are better at determining an intrinsic \lx\, which is not possible with hardness ratios or indirect methods. 

Most previous works to identify CT AGN at cosmological distances have focussed on the deepest X-ray observations in the {\it Chandra} deep field south (CDFS), where 4 Ms of {\it Chandra} data and 3Ms of \xmm\ data exist and the {\it Chandra} deep field north, where 2 Ms of {\it Chandra} data exist \citep{tozzi06, georgantopoulos09, alexander11, comastri11, brightman12b, iwasawa12, georgantopoulos13}. However, these works have so far turned up only a few tens of sources, so lack the sample size for robust statistical analyses. Due to these small numbers, only a few examples exist where CT AGN activity has been found with some connection to their host galaxies, such as the case of K20-ID5, a massive star forming galaxy hosting a CT AGN with evidence for a powerful AGN-driven outflow \citep{foersters14}, or LESS J033229.4-275619, a CT AGN associated with a compact starburting galaxy \citep{gilli14}.

Furthermore, these deep fields lack the area to find rare luminous Compton thick sources. In order to assess the cosmological significance of CT AGN and their connection to their host galaxies, a large homogeneous sample covering a wide range in luminosity and redshifts must be constructed. The goal of our work presented here is to create such a sample, using the most up to date X-ray spectral models and redshifts and obtaining the best estimate of the intrinsic \lx.

\subsection{Characterising absorption}

Much previous work has shown that the fraction of obscured AGN decreases with increasing \lx\ \citep[e.g.][]{ueda03,hasinger08,burlon11,brightman11b}. This has generally been interpreted within the framework of the so called `receding torus model' which forms the basis of a luminosity dependent unified scheme. This explains that the covering factor of the obscuring torus decreases with increasing luminosity due to the recession of the inner wall of the torus. The inner wall recedes due to the more luminous central source sublimating dust at larger radii \citep{lawrence91}. Furthermore, an increase in this fraction for constant \lx\ with increasing redshift has also been reported \citep[e.g.][]{lafranca05,treister06,hasinger08,hiroi12,merloni14}, suggesting the covering factor of the torus also increases with redshift. However, no consensus has been reached on the reason for this apparent evolution.

X-ray spectroscopy provides a unique way of investigating the nature of the obscuring material as signatures of the obscuration are imprinted on the X-ray spectrum. For \nh$>10^{23}$ \cmsq, Compton scattering becomes significant and as such the geometry of the obscuring material becomes important, as scattering can lead to radiation being reflected into the line of sight, which varies with the covering factor of the material. As the X-ray spectral torus models described above account for Compton scattering, they can be used to infer details of the obscuring material. This has been done with several local AGN using broadband {\it Suzaku} data \citep[e.g.][]{awaki09,eguchi11,tazaki11,yaqoob12,kawamuro13}, mostly to infer the covering factor of the torus. Furthermore, the use of these models has also been successful in showing that the luminosity dependent unified scheme can explain the X-ray Baldwin effect, whereby the equivalent width of the Fe K$\alpha$ line decreases with increasing \lx, due to the reduction of the torus opening angle towards higher luminosities \citep{ricci13}.

It is not only efforts using these new spectral models that have helped determine the geometry of the circum-nuclear material in AGN using X-ray spectral data. \cite{ueda07} announced the discovery of a new type of buried AGN in which the circum-nuclear material is likely to be geometrically thick with a high covering fraction. This conclusion was reached by the measurement of a very small ($<0.5\%$) scattered fraction, \fscatt. This so called ``scattered'' emission is thought to originate from the Thompson scattering of the primary X-ray photons by hot electrons within the cone of the torus, and is common in obscured AGN \citep{turner97}, or emission from circum-nuclear plasma photo-ionised by the AGN \citep{guainazzi07}. A small scattered fraction implies a small opening angle of the torus, or otherwise an under abundance of the gas responsible for the scattering. This picture is supported by \cite{noguchi10} who found that the scattered fraction correlates with the \oiii\ narrow emission line to X-ray flux ratio. As the \oiii\ emission line also originates in gas responsible for the scattering of the X-rays, it also gives an indication of the covering factor of the gas, or its abundance.

This paper is organised in the following way: we describe in Section \ref{sec_chandata} the three \chandra\ surveys used, followed by the data reduction and spectral extraction methods in Section \ref{sec_datared}. In Section \ref{sec_specanal} we describe the spectral analysis carried out on all sources, as well as the spectral models used for this work. We present our results on CT AGN and the nature of the obscuration in Section \ref{sec_results} and discuss the results along with an investigation into the spectral fitting method in Section \ref{sec_discuss}. We summarise and conclude in Section \ref{sec_conc}. We assume a flat cosmological model with $H_{\rm 0}$=70 km s$^{-1}$ Mpc$^{-1}$ and $\Omega_{\Lambda}$=0.73. For measurement uncertainties on our spectral fit parameters we present the 90\% confidence limits given two interesting parameters ($\Delta$c-stat=4.61). The results from our spectral fitting can be found in online data tables at \url{http://www.mpe.mpg.de/~mbright/data/} along with the torus models used.

\section{Chandra survey data}
\label{sec_chandata}

For our analysis, we utilise three major \chandra\ extragalactic fields which cover a range of depths and areas, and in doing so provide wide coverage of the luminosity-redshift plane. These surveys are the ultra-deep and narrow CDFS \citep{xue11}, the medium-deep, medium-width AEGIS-XD (Nandra, et al. 2014 submitted) and the shallow but wide \chandra-COSMOS \citep{elvis09}. These surveys are also the sites of extensive multiwavelength observations including HST/WFC3 observations from the CANDELS Legacy program \citep{grogin11,koekemoer11} and {\it Herschel} observations \citep{lutz11,oliver12}. Furthermore, these three fields are the focus of the {\it NuSTAR} \citep{harrison13} extragalactic survey program, which is revealing the nature of these sources above 10 keV. The {\it Chandra} survey of each field is described below.

\subsection{Chandra Deep Field South}

The CDFS currently consists of 52 observation IDs (obsIDs) with a single pointing and a total exposure of $\sim4$ Ms. It covers 465 arcmin$^2$ and reaches on axis sensitivities of $3.2\times10^{-17}$ \ergcms\ in the full band \citep[0.5-8 keV,][X11 henceforth]{xue11}. Of the 52 observations, nine were made in 2000 \citep[$\sim$1 Ms,][]{giacconi02}, 12 made in 2007 \citep[$\sim$1 Ms,][]{luo08} and a further 31 made in 2010 ($\sim$2 Ms, X11). A further approved 3 Ms is due to be added to this field, bringing the exposure to 7 Ms (PI. Brandt). The latest published source list for this field yielded 776 sources (X11), however, in order to be as uniform in our analysis as possible, we use a source list based on the same method used for the AEGIS-XD field \citep{rangel13}. A total of 569 sources were detected to a Poisson probability limit of $4\times10^{-6}$, a higher probability threshold than X11. Optical and near-infrared counterparts to the X-ray sources are presented in Hsu et al (2014, submitted), along with spectroscopic and photometric redshifts. We use spectroscopic redshifts of qualities up to 2, otherwise we use the photometric redshift. \nreda\ sources have a redshift associated with them, \nzspeca\ of which are spectroscopic redshifts. We exclude \nstarsa\ sources identified as stars in \cite{xue11} from our analysis. Fig. \ref{fig_countz} presents the X-ray spectral count distribution and redshift distributions of these sources. In section \ref{sec_nature} where we investigate trends with redshift, we do not use photometric redshifts where the peak of the probability distribution is less than 70\%, which indicates an unreliable redshift.

\subsection{AEGIS-XD}

{\it Chandra} observations of the Extended Groth Strip (EGS) began with a single pointing observation of 200 ks in 2002 \citep{nandra05}. This was followed by a further 1.4 Ms of observations in 2005 covering a contiguous strip of 2 degrees in length, bringing the total to eight pointings each with a 200 ks nominal exposure \citep[AEGIS-X, ][]{laird09}. Between 2007 and 2009 the central area of this field was imaged further to greater depth, and referred as AEGIS-XD (Nandra, et al. 2014, submitted), also part of XDEEP2 \citep{goulding12b}. AEGIS-XD consists of 72 obsIDs in three separate but contiguous pointings with a total exposure of 2.4 Ms or 800 ks per pointing. It covers 871 arcmin$^2$ and reaches on axis sensitivities of $1.7\times10^{-16}$ \ergcms\ in the full band (0.5-8 keV). \nredb\ sources have a redshift associated with them, \nzspecb\ of which are spectroscopic redshifts. We exclude \nstarsb\ sources identified as stars from our analysis. Fig. \ref{fig_countz} presents the X-ray spectral count distribution and redshift distributions of these sources. As for CDFS, in section \ref{sec_nature} we do not use photometric redshifts where the peak of the probability distribution is less than 70\%.

\subsection{C-COSMOS}

The \chandra-COSMOS survey consists of 49 obsIDs with a total exposure of 1.8 Ms over 36 individual pointings which are heavily overlapped (80 ks per pointing, 160 ks with overlaps). All data in this survey was taken between 2006 and 2007. The total survey area covers 0.9 deg$^{2}$, whereas the central area covers 0.5 deg$^{2}$ (1800 arcmin$^2$). This reaches a sensitivity of $5.7\times10^{-16}$\ergcms\ in the full band (0.5-8 keV). This field has recently been extended to cover the full 2 deg$^{2}$ of the COSMOS survey, referred to as the \chandra-COSMOS Legacy survey (PI. Civano). For C-COSMOS, we use the source catalogue of \cite{elvis09} which yields 1761 sources. Optical identifications and redshifts for these sources were presented in \cite{civano12}, with \nredc\ sources having a redshift associated, \nzspecc\ of which were spectroscopic. We exclude \nstarsc\ sources identified as stars in \cite{civano12} from our analysis. Fig. \ref{fig_countz} presents the X-ray spectral count distribution and redshift distributions of these sources. In section \ref{sec_nature}, we do not use photometric redshifts where a second peak in the probability distribution is present and the primary peak is less than 70\% as presented in \cite{salvato09, salvato11}.

\subsection{Combined sample}

\begin{figure}
\begin{center}
\includegraphics[width=90mm]{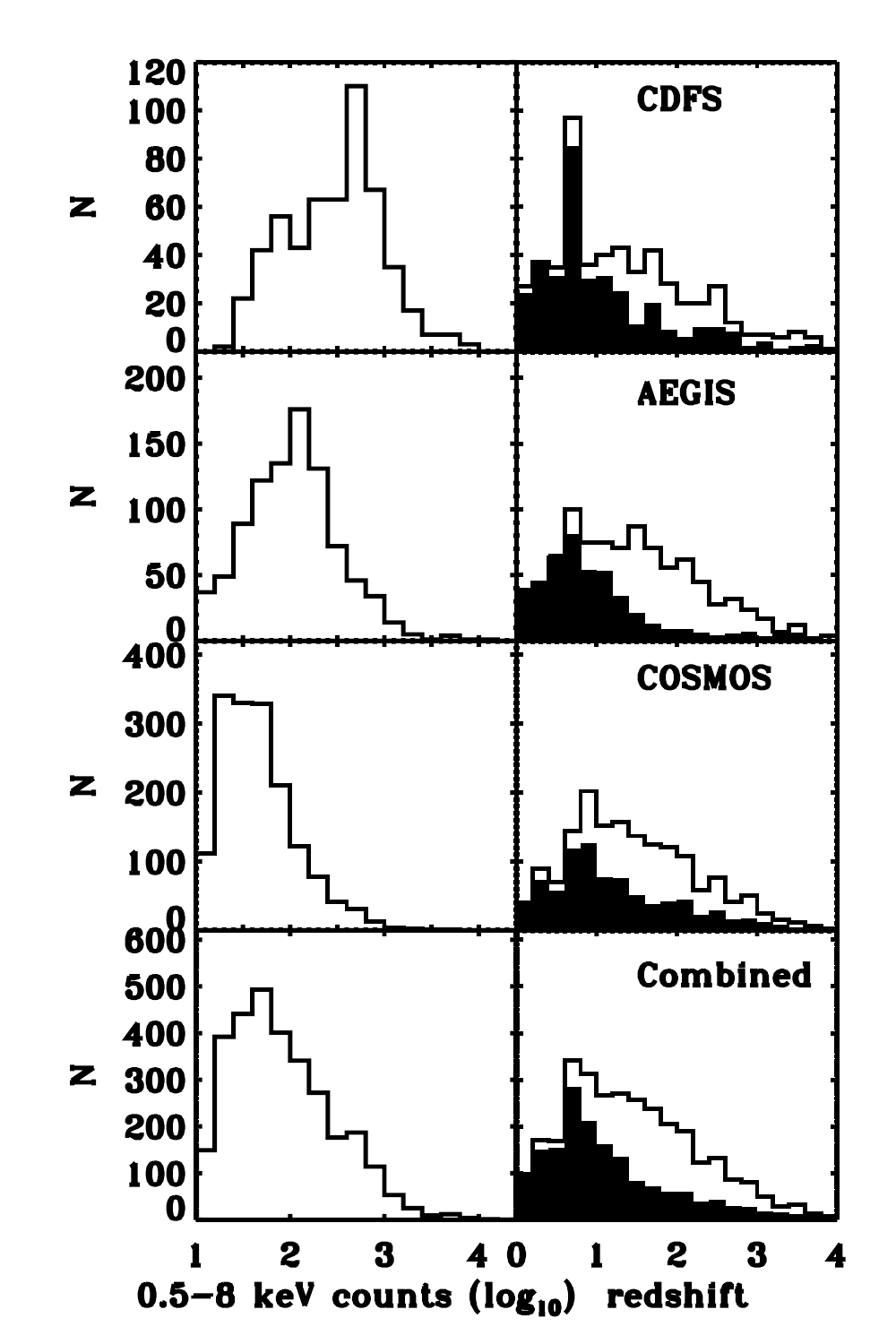}
 \caption{Distributions of spectral counts for the individual fields and the combined sample (left panels) and redshift distributions (right), where the solid histograms represent the spectroscopic redshift distributions. While a few sources beyond z=4 (the range of the plot) exist in these surveys, their numbers are very small relative to the rest of the sample.}
 \label{fig_countz}
 \end{center}
\end{figure}

When combining the three \chandra\ surveys above, we have a total of \nsrc\ extragalactic sources (having removed sources identified as stars), \nred\ of which have a redshift associated, giving a redshift completeness of \zcomp\%. We neglect those sources without redshifts in this work. Due to the high level of completeness, we do not expect this to introduce any bias into our work. We summarise the basic properties of these surveys, and their combined properties in Table \ref{survey_table}. We also present the combined spectral count and redshift distributions are in Fig. \ref{fig_countz}. 

\begin{table*}
\centering
\caption{Summary of the basic properties of the three \chandra\ surveys used in this work. Column (1) gives the survey name, column (2) gives the area of the survey in armin$^{2}$, column (3) gives the sensitivity of the survey in the 0.5-8 keV band, in units of \ergcms, column (4) gives the total number of non-stellar sources detected in the survey, column (5) gives the number of redshift identifications made for these sources and column (6) gives the redshift completeness. }
\label{survey_table}
\begin{center}
\begin{tabular}{l r c r r c}
\hline
Survey & Area & Sensitivity & non-stellar sources & redshift IDs & redshift completeness \\
(1) & (2) & (3) & (4) & (5) & (6) \\
\hline
CDFS & 465 & $3.2\times10^{-17}$ & \nsrca\ & \nreda\ & \zcompa\% \\
AEGIS & 871 & $1.7\times10^{-16}$ & \nsrcb\ & \nredb\ & \zcompb\% \\
COSMOS & 3240 & $5.7\times10^{-16}$ & \nsrcc\ & \nredc\ & \zcompc\% \\
Combined & 4576 & - & \nsrc\ & \nred\ & \zcomp\% \\

\hline
\end{tabular}
\end{center}
\end{table*}

\section{Data reduction and spectral extraction}
\label{sec_datared}

We reduce the \chandra\ survey data in a uniform manner, screening for hot pixels and cosmic afterglows as described in \cite{laird09} using the {\sc ciao} data analysis software version 4.2. Individual source spectra were extracted from these processed data using the {\sc acis extract} (AE) software version 2011-03-06 \citep{broos10}.  \footnote{The {\sc acis extract} software package and User's Guide are available at http://www.astro.psu.edu/xray/acis/acis\_analysis.html}. AE extracts spectral information for each source from each individual obsID based on the shape of the local point spread function (PSF) for that particular position on the detector. We choose to use regions where 90\% of the PSF has been enclosed at 1.5 keV. Background spectra are extracted from an events list which has been masked of all detected point sources in that particular field, using regions which contain at least 100 counts. AE also constructs response matrix files (RMF) and auxiliary matrix files (ARF). The data from each obsID are then merged to create a single source spectrum, background spectrum, RMF and ARF for each source. Histograms of the background-subtracted source counts are presents for each field individually and for the combined sample in Fig \ref{fig_countz}.

\section{X-ray spectral analysis}
\label{sec_specanal}

We take a uniform approach to the spectral analysis of all \nred\ sources in our combined sample of the three surveys. The method we use is a modification of that described in BU12, which was shown to be effective in identifying Compton thick sources in the CDFS. The main modifications made are those that account for the lower signal to noise spectra from the AEGIS and COSMOS surveys. We carry out spectral fitting using {\sc xspec} version 12.6.0q. As the number of spectral counts available for spectral fitting is generally low in our sample, with the majority of the sample having less than 100 counts, we do not wish to significantly group the spectra so as to preserve spectral features. Therefore, we only lightly group the spectra with a minimum of one count per bin using the {\sc heasarc} tool {\tt grppha}. For the spectral fit statistic, we use the Cash statistic \citep[c-stat,][]{cash79} which uses a Poisson likelihood function and is hence most suitable for low numbers of counts per bin. Furthermore, \cite{tozzi06} showed using simulations that the Cash statistic recovers spectral parameters more reliably than the commonly used $\chi^2$ statistic in the low count regime. \cite{lanzuisi13} also conducted an investigation into the use of c-stat with respect to $\chi^2$ for C-COSMOS sources, finding that c-stat gives a 30\% lower error on the measured parameters. Only channels in the observed frame 0.5-8 keV range are used due to the decline in the effective area of \chandra\ ACIS-I outside this range. Due to the wide range of redshifts in our sample, this observed frame wavelength range probes a large range of rest-frame energies. We address this potential biasing effect further on in our analysis.

\subsection{X-ray spectral models}
\label{sec_specmo}

As the main goal of this work is to identify and characterise the most heavily obscured sources, including those which are Compton thick, we use up-to-date spectral models. We use the models presented in \cite{brightman11}, which employ Monte-Carlo simulations to account for Compton scattering and the geometry of the obscuring material, both of which are important for heavily obscured sources. They also include self consistent iron K$\alpha$ emission and describe spherical and torus distributions of circumnuclear material. As in BU12, rather than attempting to constrain the torus opening angle from the spectrum, three different cases where torus opening angles were fixed at 60\degree, 30\degree\ and 0\degree\ were tested, where 0\degree\ is essentially a 4$\pi$ spherical distribution. For opening angles $>0$\degree\ we include a secondary power-law component in the fit, which represents intrinsic scattered emission, reflected by hot electrons filling the cone of the torus. This component may also represent host galaxy emission for low luminosity sources, or thermal plasma emission, also thought to originate in the cone of the torus. In any case, the level of this emission with respect to the intrinsic emission is of order a few per cent. From colour-colour analysis of CDFN sources, BN12 showed that this spectral complexity is very common in high redshift sources, as it is in the local universe, and therefore it is essential to account for it. The model combinations used are as follows:

\begin{itemize}

\item{A) The torus model of BN11 with a fixed opening angle of 60$^\circ$, and edge on orientation, accompanied by a scattered power-law, with $\Gamma_{\rm scatt}$ fixed to the value of the primary power-law. This model has four free parameters, \nh; $\Gamma$; and the normalisations of the power-laws, A$_1$ and A$_2$}
\item{B) The torus model of BN11 with a fixed opening angle of 30$^\circ$, and edge on orientation, accompanied by a scattered power-law, with $\Gamma$ fixed to the value of the primary power-law. This model has four free parameters, \nh; $\Gamma$; A$_1$ and A$_2$}
\item{C) The spherical model of BN11, a scenario in which the X-ray source is completely covered with 4$\pi$ steradians of obscuring material. This model has three free parameters,  \nh; $\Gamma$ and A$_1$}. We include no scattered component for this model as it represents the case where there is no escape route for the primary radiation to be scattered into the line of sight. \item{D) simple power-law model, with two free parameters, the power-law index, $\Gamma$ and the normalisation, A$_1$, representing unobscured X-ray emission.}
\end{itemize}

\begin{figure*}
\includegraphics[width=180mm]{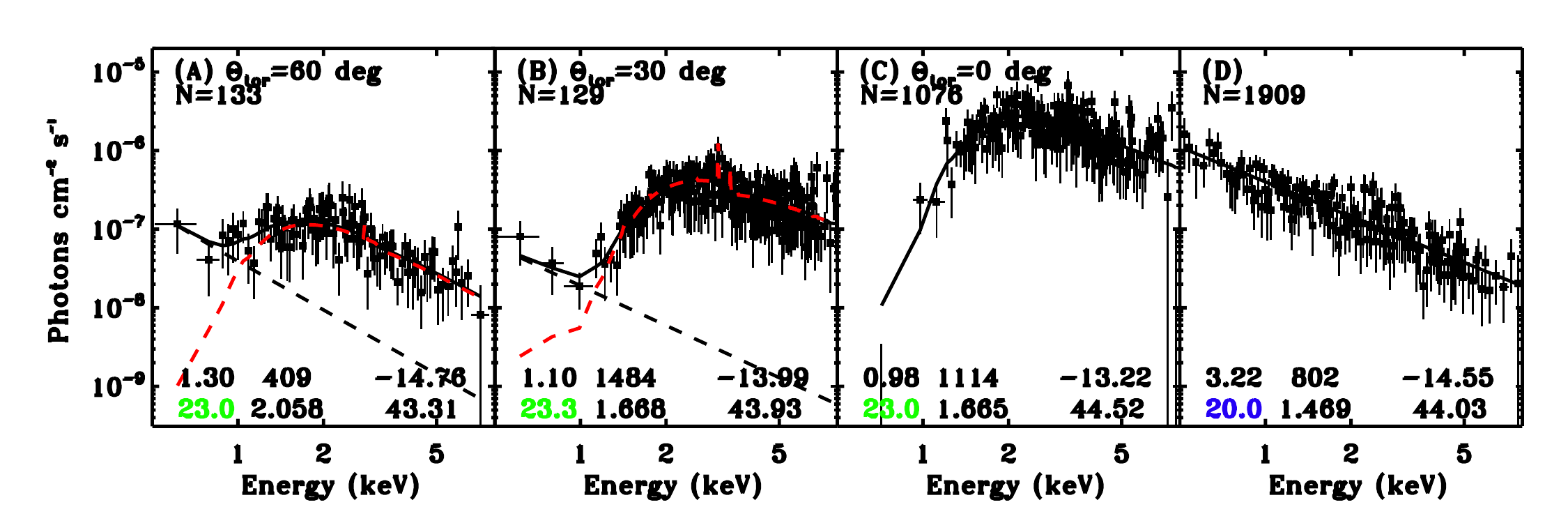}
\caption{Four example spectra from CDFS, one for each model combination as described in Section \ref{sec_specmo}. Red dashed lines show the torus model which represents the heavily obscured primary emission, whereas dashed black lines show the secondary power-law component, representing intrinsic scattered emission. Solid black lines show the total model spectrum. From left to right, the first three panels show the torus model with 60\degree, 30\degree\ and 0\degree\ opening angles. Note the relatively stronger scattered component in the 60\degree\ case, and the absence of this emission in the 0\degree\ case. This supports the picture that the larger the opening angle, the stronger the soft scattered emission. The total number of spectra best fit by these models is given at the top of each panel, whereas in the bottom of each panel, from top-left to bottom-right, are the redshift, spectral counts, log$_{10}F_{\rm X}$ (observed 0.5-8 keV, \ergcms), log$_{10}$\nh, $\Gamma$ and log$_{10}$\lx\ (intrinsic 2-10 keV, \ergs). The spectra have been rebinned for plotting purposes. }
\label{fig_mospec}
\end{figure*}

We show an example spectrum for each model combination in Fig. \ref{fig_mospec} from the CDFS sample. 

\subsection{X-ray spectral fitting}

\subsubsection{Model selection}
\label{sec_modsel}

Each of the four model combinations are fitted to the source spectrum in turn with at least 100 iterations with a critical delta of $1\times10^{-5}$. The best fit model combination is chosen to be that which presents the lowest c-stat value. However, the above list of model combinations have different numbers of free parameters, and thus we must penalise more complex models to ensure that each free parameter/spectral component is statistically required. We do this by requiring that for each extra free parameter a model combination has, it must present an improvement of $\Delta$c-stat$>2.71$ over the model combination without that free parameter.  For example, for model C, with three free parameters to be chosen over a simple power-law (model D), with only two free parameters, the c-stat improvement of model C over model D must be at least 2.71. This criterion is roughly consistent with the 90\% confidence level using an F-test, and was used by \cite{tozzi06} in the same way in their spectral analysis of CDFS sources.  It is high enough so that when identifying heavily obscured sources, the rate of contamination is low, but not so high as to be too conservative. While this method is strictly only true for nested models, where one model is a special case of another, we find it fits our purpose here. Indeed, as \cite{buchner14} (henceforth B14) point out, this criterion is actually consistent with the Akaike Information Criterion (AIC), which is based in information theory. Here the over-fitting penalisation is just twice the difference in parameters (i.e. $\Delta$c-stat$>2.0$).

We conduct simulations to investigate the use of this criterion. For three sources, aegis\_540, aegis\_236 and aegis\_228 with z$\simeq$0.5 and \nh=2.2$\times10^{24}$ (model B), 7$\times10^{22}$ (model C) and 0 (model D, unobscured) \cmsq\ respectively, we simulate 100 spectra each in three different count regimes, 10-30, 30-100, and 100-300 counts. We fit the simulated spectra in the same way as above, and define the confidence levels for each model selection as the number of spectra where the best fit model is the correct one as a fraction of the number of all spectra where the model is chosen to be the best fit model. Here we treat model A and B as the same, as they have the same number of free parameters, with the only difference being the opening angle of the torus. Table \ref{sims_table} presents these confidence levels, showing that with more than 30 counts in the spectrum, our choice of $\Delta$c-stat is indeed consistent with $>$90\% confidence.

\begin{table}
\centering
\caption{Table presenting the confidence levels for the model selection, given the choice of $\Delta$cstat=2.71 for three different count regimes, defined as the fraction of spectra where the best fit model is the correct one, from simulations.}
\label{sims_table}
\begin{center}
\begin{tabular}{l r r r}
\hline
counts & A/B & C & D \\
(1) & (2) & (3) & (4) \\
\hline
10-30     &  0.88&  0.67&  0.73\\
30-100    &  0.86&  0.95&  0.96\\
100-300    &  0.91&  0.94&  1.00\\
\hline
\end{tabular}
\end{center}
\end{table}

Furthermore, the case of a fixed power-law index is explored. It was shown in BU12 that for less than 600 counts the constraint on $\Gamma$ is poor ($\Delta\Gamma>$0.5). This introduces large uncertainties into the spectral fit which can be reduced by fixing $\Gamma$ in the fit. We do this for sources with less than 600 counts in the spectrum. For sources with more than 600 counts, we find that the mean spectral index of our sample is \meangamml, thus we use a fixed value of $\Gamma=1.7$ for sources with less than 600 counts. We do however also allow a consideration for intrinsically steep or flat spectra. If the best fit model for sources with less than 600 counts, where $\Gamma$ is free is a significantly better fit than the best fit model where $\Gamma$ is fixed, using the criterion of $\Delta$c-stat$>2.71$, we choose the model with $\Gamma$ free as the best fit model. For CDFS, \nhicntsa\ sources have more than 600 counts. Of the remaining \nlocntsa\ sources, \nbigcdifa\ spectral fits require that $\Gamma$ be left free. For AEGIS, \nhicntsb\ sources have more than 600 counts, and of the remaining \nlocntsb\ sources, \nbigcdifb\ spectral fits require that $\Gamma$ be left free. Finally for COSMOS, only \nhicntsc\ sources are above the count limit, whilst from \nlocntsc\ of the remaining sources, \nbigcdifc\ also require that $\Gamma$ be left free.

\subsubsection{Reducing contamination}

The main goal of this work is to identify and characterise the most heavily obscured Compton thick (\nh$>10^{24}$ \cmsq) sources. In BU12, a low $\Delta$c-stat criterion of 1.0 was used to pick out Compton thick sources in the CDFS and assess the evolution of the Compton thick fraction, using simulations to statistically correct for contamination and incompleteness. Here we aim to present the cleanest, most reliable sample of Compton thick sources so that the population may be characterised on a case by case basis. Also, as AEGIS and COSMOS have lower signal to noise spectra, the contamination rate will be higher when analysing these fields. For this reason the above $\Delta$c-stat criterion of 2.71 is used across all three fields. The majority of the sources contaminating the Compton thick sample are in fact bright unobscured sources, where the spectral fit includes a strong scattered component with a weak underlying torus component, which may be fit to a statistical hard fluctuation. While this may also represent some partial covering scenario \citep[e.g.][]{mayo13}, it is more likely to be falsely identified absorption.  We therefore place an upper limit of 10\% on the scattered component for selecting heavily absorbed sources. In addition we also note that for 0.5-8 keV fluxes greater than $10^{-14}$ \ergcms, the fraction of Compton thick sources is negligible \citep[BU12]{gilli07,akylas12}, and hence we also place an upper limit of $10^{-14}$ \ergcms\ for Compton thick sources. For CDFS we exclude \nhifsca\ sources with high \fscatt\ values as being heavily obscured, and instead select the simple power-law model, whereas we exclude \nhifscb\ in AEGIS and \nhifscc\ in COSMOS. As for sources with fluxes greater than $10^{-14}$ \ergcms\ which we exclude as being Compton thick, there are \ncthifxb\ in AEGIS and \nhifscc\ is COSMOS. We make no such exclusions in CDFS.

\subsubsection{Identifying heavily absorbed sources not found through spectral fitting}

\begin{figure}
\includegraphics[width=80mm]{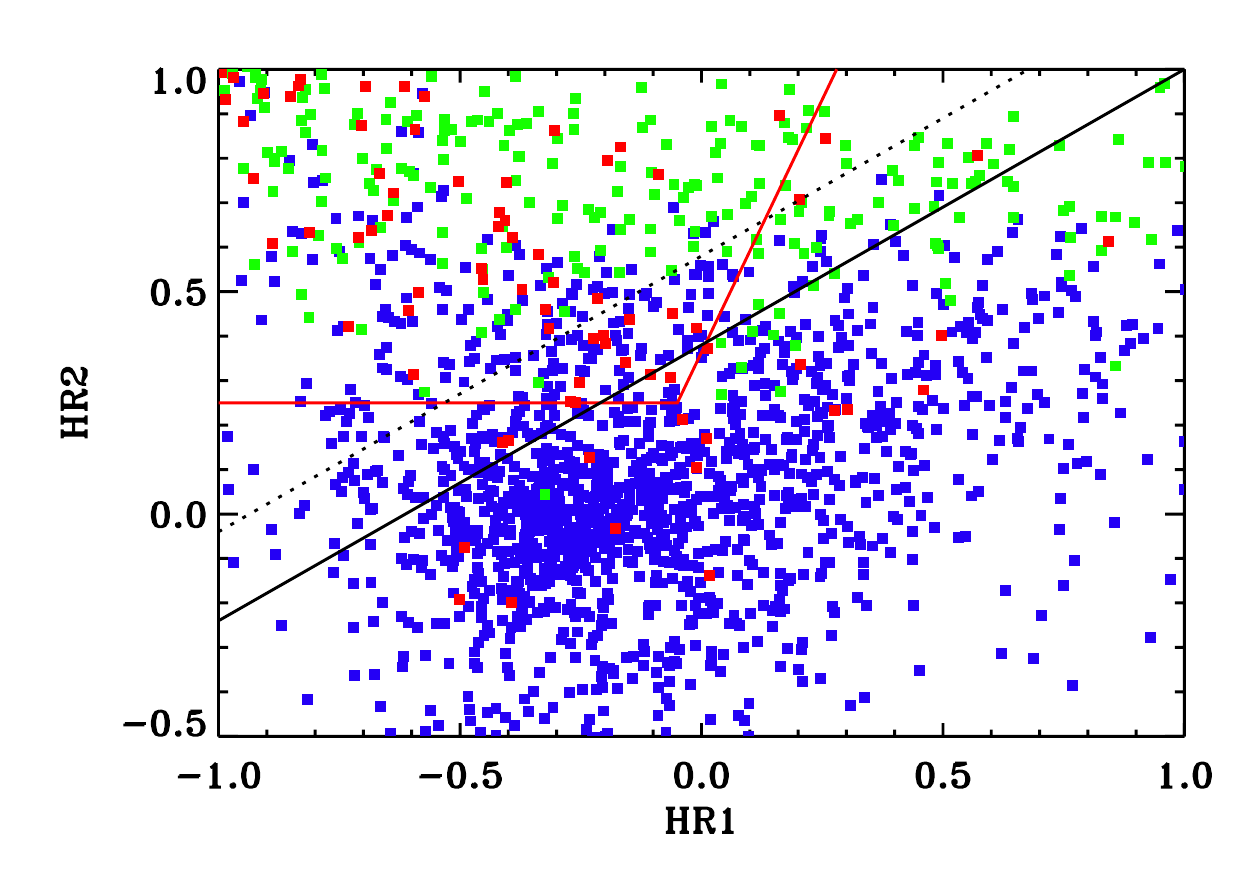}
\caption{X-ray colour-colour scheme from BN12 for all sources in our sample up to z=2. Data points are colour coded by \nh, where blue points have \nh$<10^{23}$ \cmsq, green points have $10^{23}<$\nh$<10^{24}$ \cmsq\ and red points have \nh$>10^{24}$ \cmsq. The solid black line marks the BN12 selection line for \nh$<10^{23}$ \cmsq\ sources. We identify heavily obscured sources using the more conservative dotted line. The solid red line is the selection wedge for CT AGN from BN12.}
\label{fig_hrplot}
\end{figure}

In order to maximise the number of heavily absorbed AGN identified here, we use colour-colour analysis to pick out sources which are likely to be heavily absorbed, but that have not been identified through the above method. Simple hardness ratios are commonly used to identify heavily absorbed sources. \cite{brightman12} however showed that these can often be unreliable as soft excess emission can make a heavily absorbed source look soft. BN12 showed that using a combination of two hardness ratios, one of which isolates the soft emission, can be very effective at identifying these sources. The rest-frame bands used in this are 1-2 keV (SB1), 2-4 keV (SB2) and 4-16 keV (HB) and the hardness ratios are defined as HR1=(SB2-SB1)/(SB2+SB1) and HR2=(HB-SB2)/(HB+SB2) using count rates converted into fluxes. Heavily absorbed sources are found in the region where ${\rm HR2 > 0.62 \times HR1 + 0.38}$. Due to their rest-frame definitions, these hardness ratios are best suited to sources at z=1 for \chandra\ data. We examine all sources in our analysis where the best fit model is the simple power-law with no absorption (model D above) and where $\Gamma<1$, which typically indicates an unidentified heavily absorbed source. For z$<$2, if ${\rm HR2 > 0.62 \times HR1 + 0.38}$ or for z$>$2 (where HR1 is out of the \chandra\ bandpass), if HR2$>$0.5, an unobscured source is very unlikely, so we redetermine the best fit model from A, B or C as described above. These models all include the \nh\ parameter. In this way, we find an additional \nhrabsa\ heavily obscured AGN (\nh$>10^{23}$ \cmsq) in the CDFS, \nhrabsb\ heavily obscured AGN (\nh$>10^{23}$ \cmsq) in AEGIS and \nhrabsc\ heavily obscured AGN in COSMOS. Figure \ref{fig_hrplot} shows the BN12 scheme and its selection lines, with data from our sample plotted, once the above additions were made.

BN12 further suggested a selection wedge to specifically select sources with \nh$>10^{24}$ \cmsq, shown as a solid red line in Figure \ref{fig_hrplot}. We find that 78/105 (74\%) of the sources with best-fit \nh$>10^{24}$ \cmsq\ in this work at z$<$2 lie within this area.

\section{results}
\label{sec_results}

\subsection{X-ray spectral parameters}

We present the spectral fit parameters for all \nred\ sources we have analysed here. The data can be found in online data tables at \url{http://www.mpe.mpg.de/~mbright/data/} along with the torus models used here. We summarise the results with the average parameters for each best fit model in Table \ref{tab_models}.

\begin{table*}
\centering
\caption{Details of the spectral fitting results by best fit model. Column (1) gives the name of the model described in section \ref{sec_specmo}, column (2) gives the number of spectra best fit by this model, column (3) gives the number of spectra where \nh$>10^{23}$ \cmsq, best fit by each model, column (4) gives the number of spectra with more than 600 counts in the spectrum, and where \nh$>10^{23}$ \cmsq, best fit by each model, column (5) gives the mean $\Gamma$ and standard error for these spectra of column (4) and column (6) gives the average \fscatt\ and standard error of the spectra in column (4) }
\label{tab_models}
\begin{center}
\begin{tabular}{l r r r r r}
\hline
Model & All & \nh$>10^{23}$ \cmsq\ & \nh$>10^{23}$ \cmsq\ & $<\Gamma>$ & $<$\fscatt$>$ \\
 & & & $>$600 counts \\
(1) & (2) & (3) & (4) & (5) & (6)  \\
\hline
(A) 60\degree\ torus &   131 (  4.0\%) &   127 ( 16.0\%) &     9 ( 28.0\%) &  1.59$\pm$ 0.14&  7.24$\pm$ 2.38\\
(B) 30\degree\ torus &   134 (  4.1\%) &   127 ( 16.0\%) &     6 ( 18.0\%) &  1.72$\pm$ 0.15&  2.91$\pm$ 0.93\\
(C) 0\degree\ torus &  1085 ( 33.4\%) &   506 ( 66.0\%) &    17 ( 53.0\%) &  1.70$\pm$ 0.08& - \\
(D) power-law &  1897 ( 58.4\%) & - &  - & - & - \\
\hline
\end{tabular}
\end{center}
\end{table*}

\begin{figure*}
\includegraphics[width=180mm]{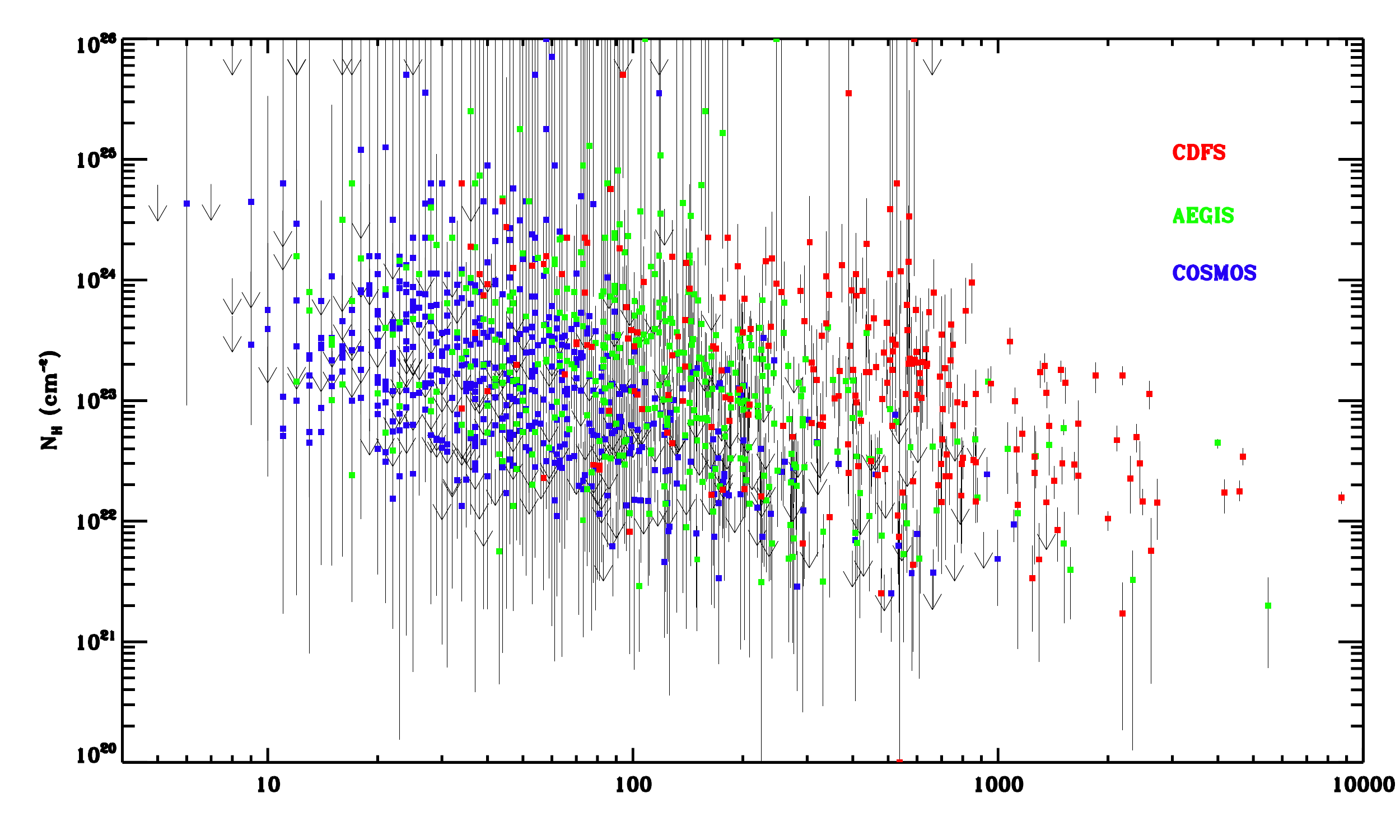}
\caption{Measurements of \nh, showing the level of uncertainty (90\% confidence level) achieved as a function of 0.5-8 keV counts for sources where the \nh\ has been constrained to be greater than zero. The data points are colour-coded by survey. Upper limits are shown as arrows. }
\label{fig_ctsnh}
\end{figure*}

\begin{figure*}
\includegraphics[width=180mm]{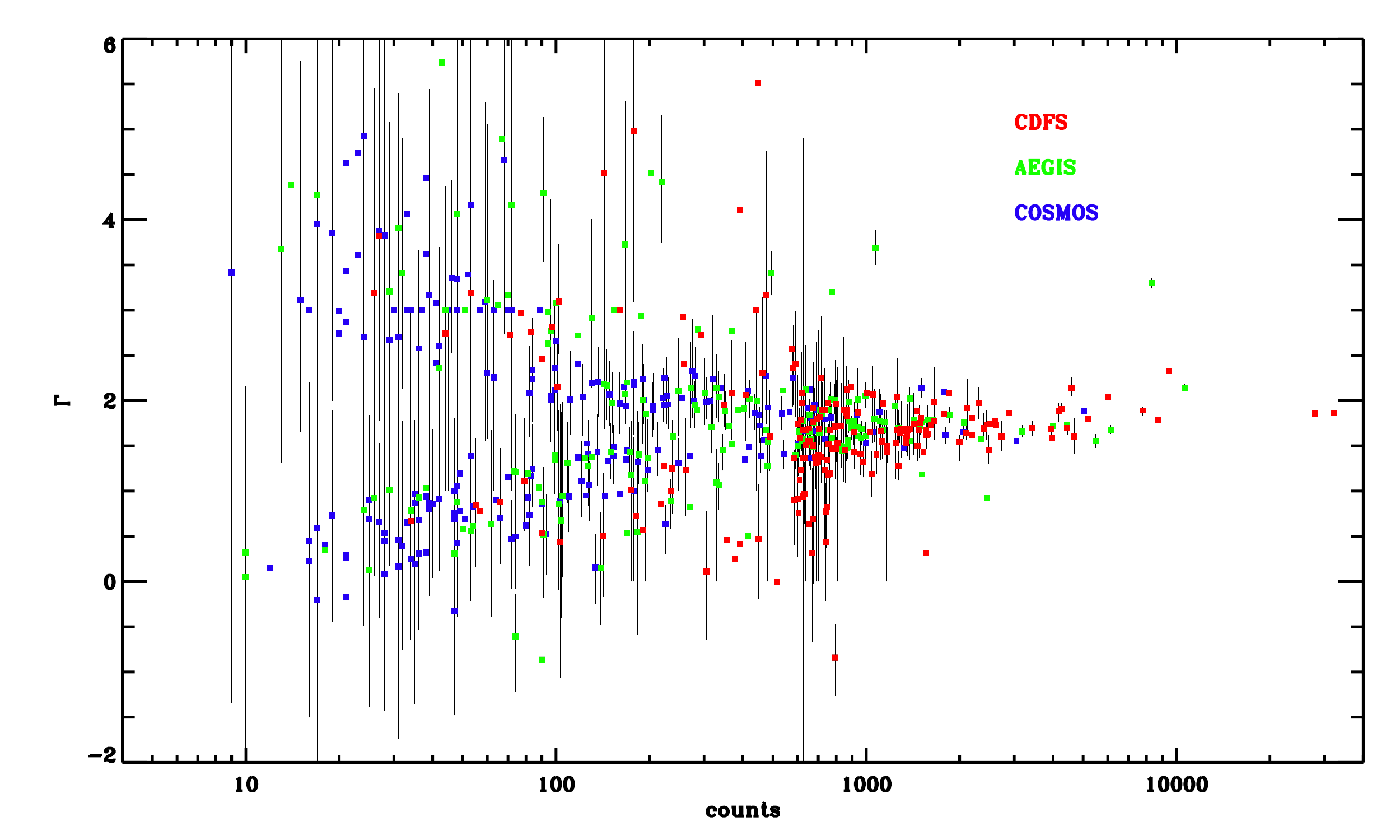}
\caption{Measurements of $\Gamma$ showing the level of uncertainty (90\% confidence level) achieved as a function of 0.5-8 keV counts. The data points are colour-coded by survey. If $\Gamma$ has been fixed in the spectral fit, it has not been plotted. }
\label{fig_ctsgam}
\end{figure*}

\begin{figure}
\includegraphics[width=90mm]{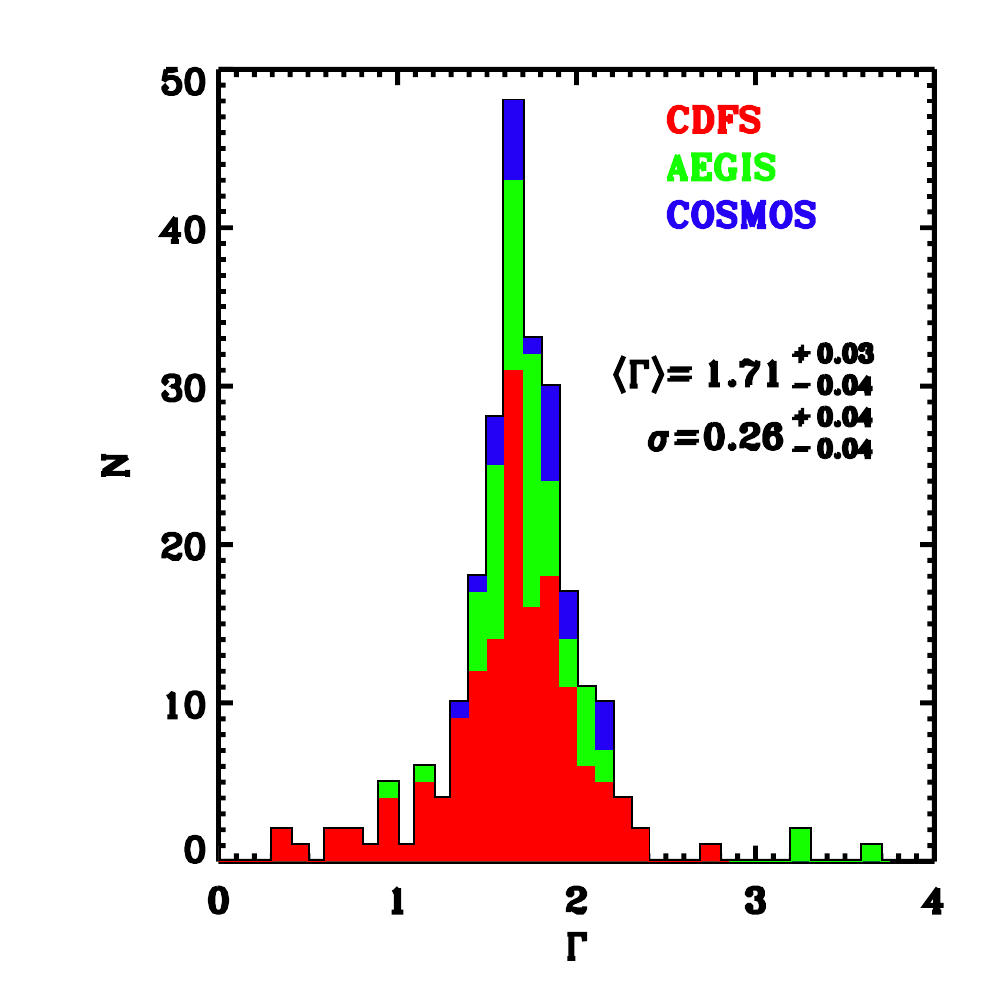}
\caption{Distribution of the photon index, $\Gamma$ for all sources with more than 600 spectral counts. The mean of the distribution is \meangamml\ with a standard deviation of \sdgamml, calculated from the maximum likelihood method of Maccacaro et al (1988)}
\label{fig_gamdist}
\end{figure}

\begin{figure}
\includegraphics[width=90mm]{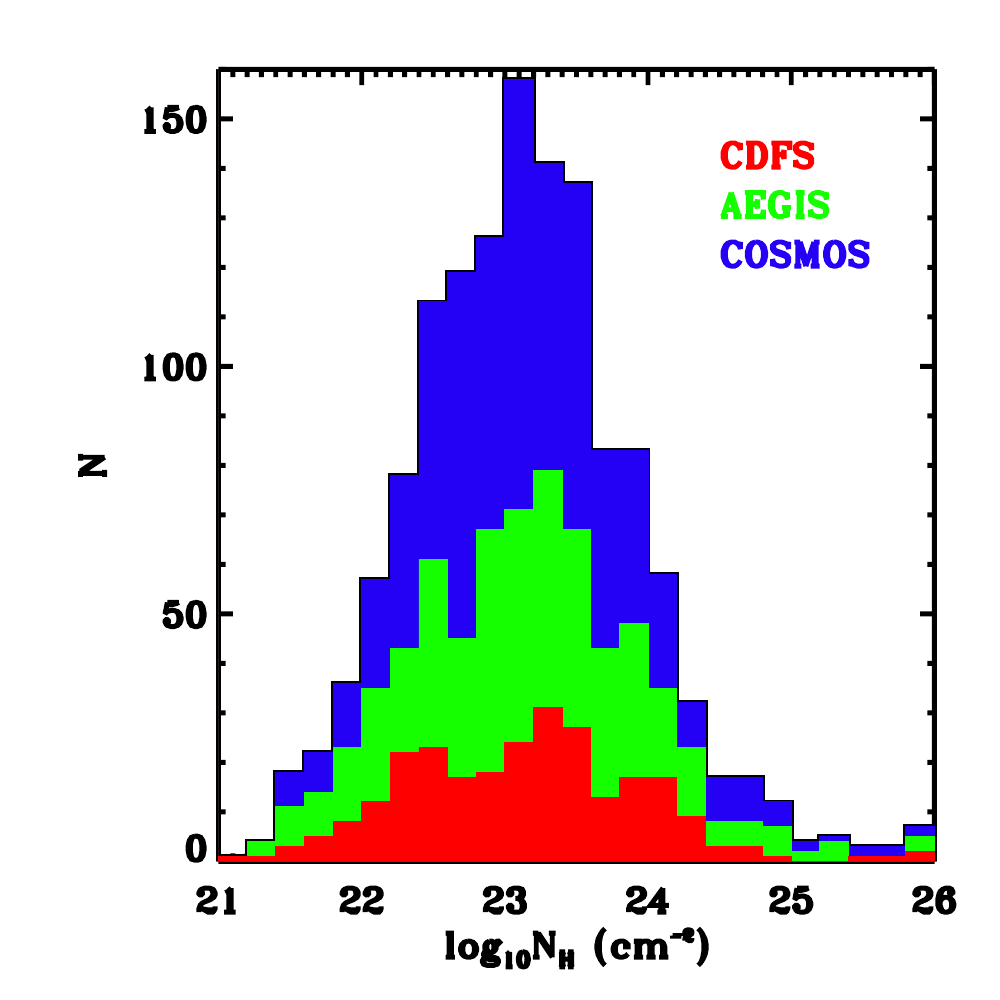}
\caption{Observed distribution of the column density, \nh\ for all sources with measured absorption. We do not show sources with no measured absorption, but they number \nunobs\ in total, \nunobsa\ in CDFS, \nunobsb\ in AEGIS and \nunobsc\ in COSMOS.}
\label{fig_nhdist}
\end{figure}

Two important parameters which can be measured directly from the X-ray spectrum are the line of sight column density, \nh, and the photon index, $\Gamma$. Figures \ref{fig_ctsnh} and \ref{fig_ctsgam} show the measurements of these parameters, along with the level of uncertainty we achieve, as a function of the number of counts in the spectrum. As described above, for less than 600 counts, $\Gamma$ is fixed at 1.7 unless a free $\Gamma$ gives a significantly better fit. The intrinsic distributions of these parameters are important in AGN studies, however this requires careful treatment of many effects which is outside of the scope of this paper, where we are focussing on identifying and characterising the most obscured AGN. We do however present the observed distributions of these parameters in Fig. \ref{fig_gamdist} and Fig. \ref{fig_nhdist}. 

\begin{figure}
\includegraphics[width=90mm]{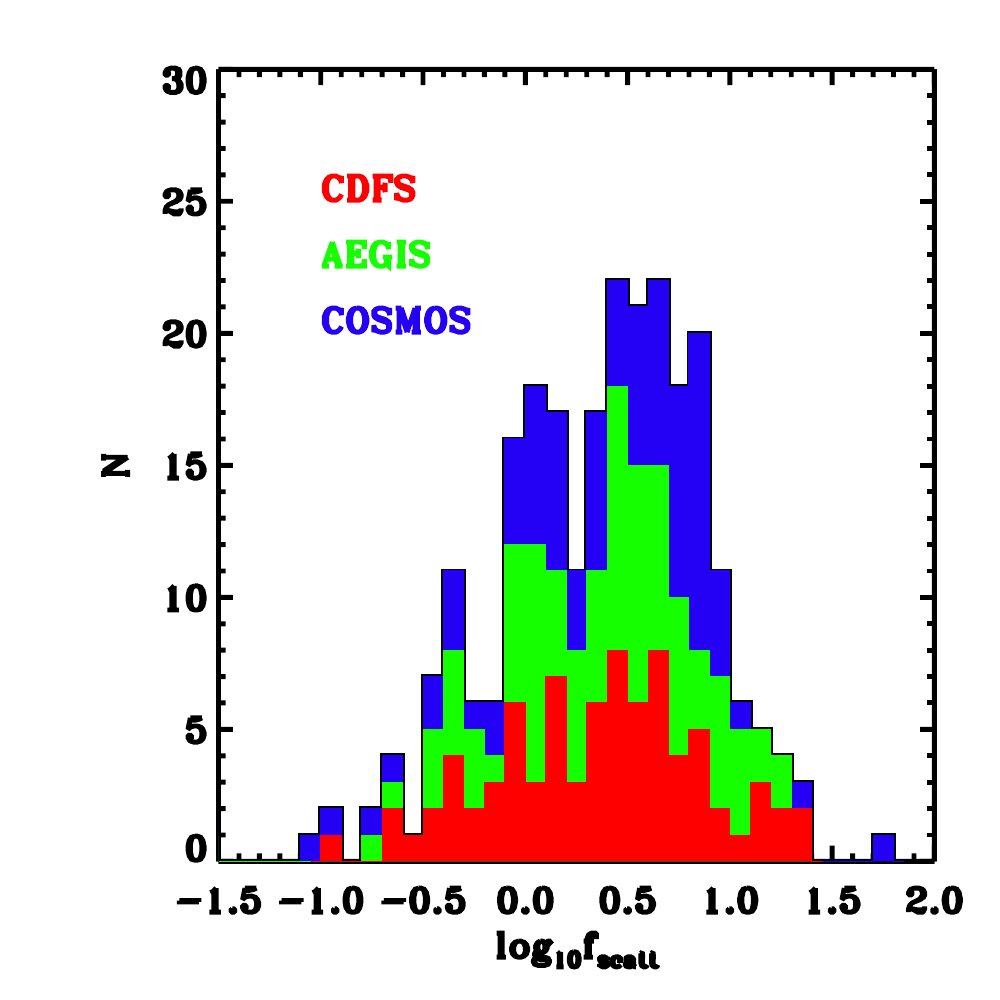}
\caption{Distribution of the \fscatt\ parameter, the relative strength of the soft excess component of the absorbed sources, shown here as a logarithmic per cent. An upper limit of \fscatt=10\% is applied when selecting the best fit model, however larger values of \fscatt\ are permitted when X-ray colours indicate strong soft emission combined with a hard component. The logarithmic mean of this distribution is \meanfscatt\%}
\label{fig_fscattdist}
\end{figure}

A further parameter which can be directly measured in the X-ray spectrum is the strength of the scattered emission, where it is present. This is parameterised as \fscatt, which is the ratio of the normalisation of the secondary power-law used to fit the soft excess, to the intrinsic normalisation of the primary absorbed component modelled here by the torus model. Both are defined at rest-frame 1 keV. \fscatt\ gives information about the filling factor and optical depth of the scattering medium and hence, the covering factor of the torus. Fig. \ref{fig_fscattdist} shows the distribution of \fscatt\ in this sample.

\begin{figure*}
\includegraphics[width=180mm]{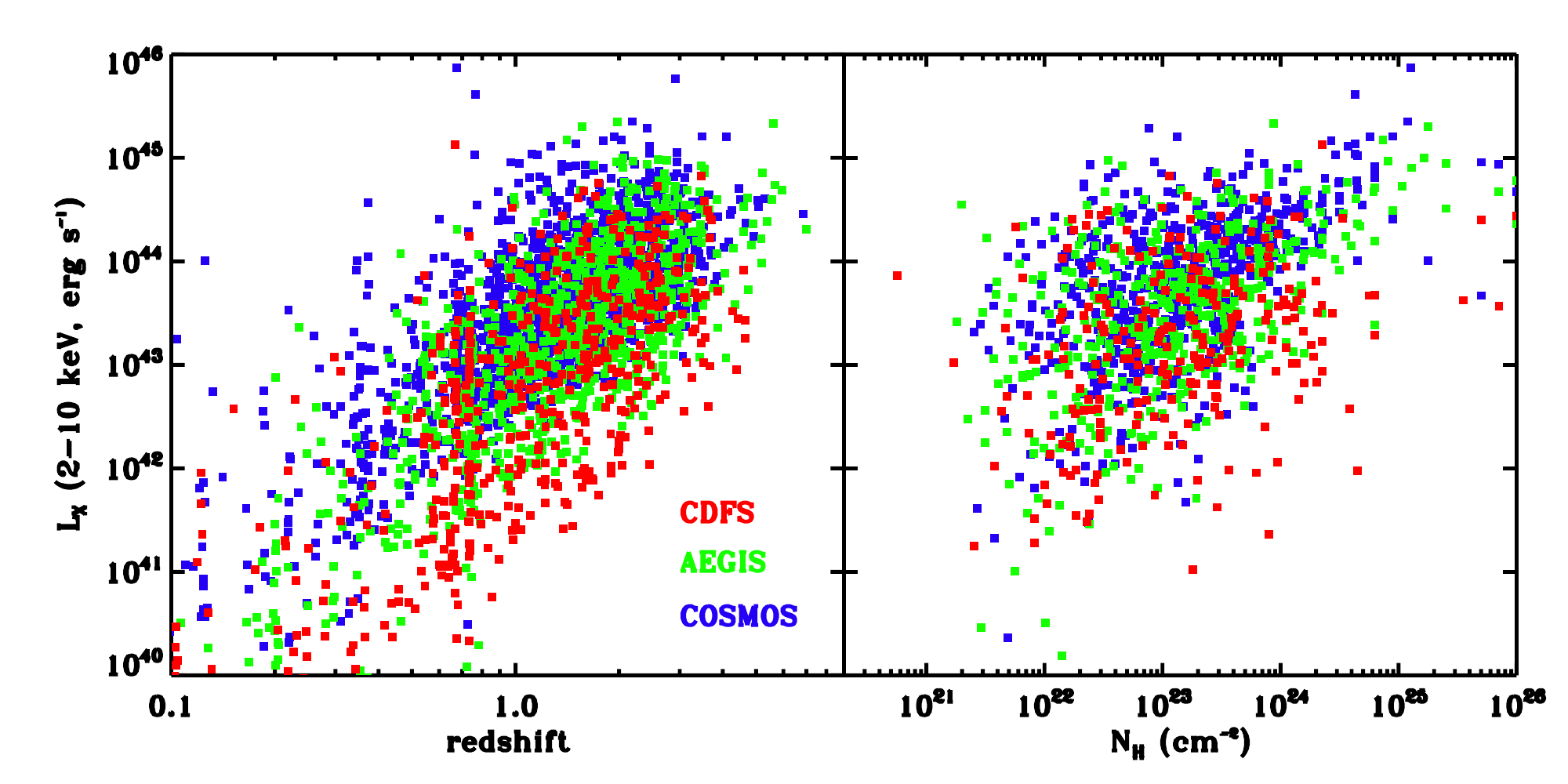}
\caption{Relative coverage of the luminosity-redshift and luminosity-\nh\ planes for the three surveys used. X-ray luminosities are intrinsic rest-frame 2-10 keV luminosities having been corrected for absorption.}
\label{fig_zlxnh}
\end{figure*}

\begin{figure}
\includegraphics[width=90mm]{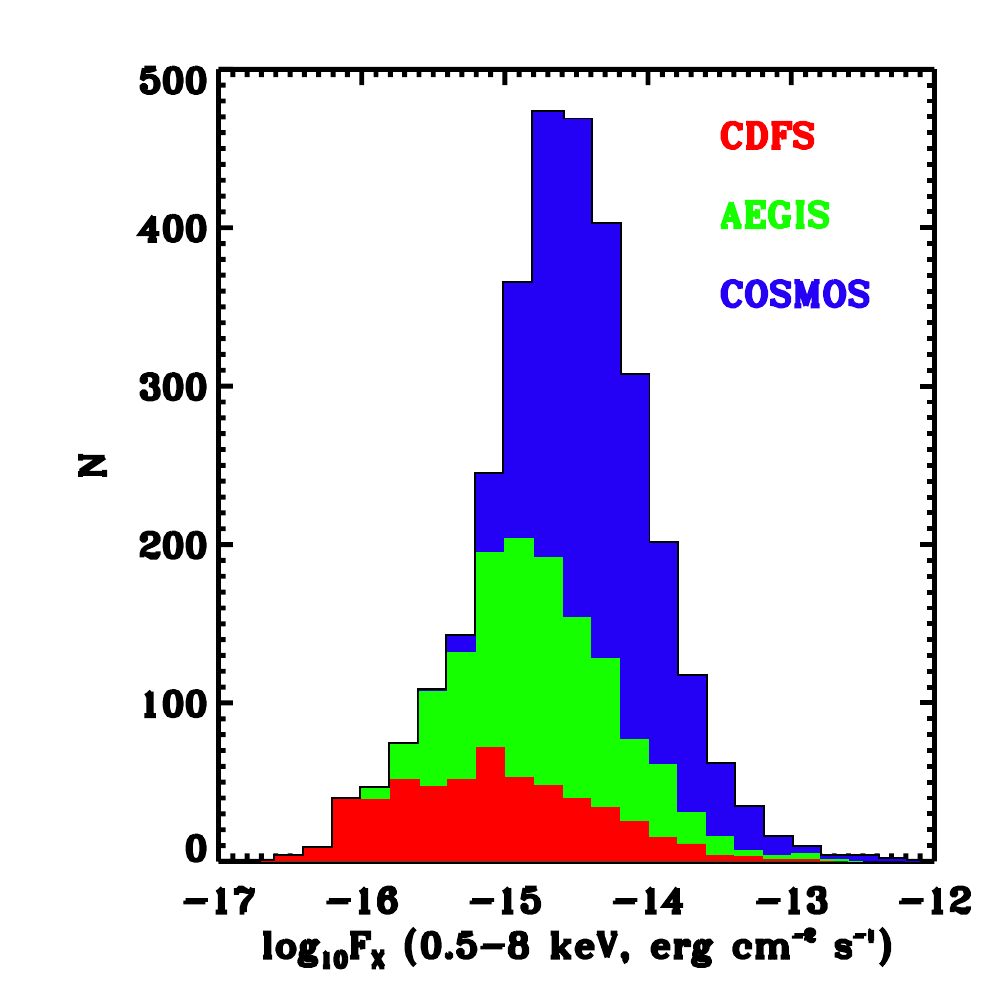}
\caption{Distribution of the observed 0.5-8 keV fluxes for all sources in the sample}
\label{fig_fxdist}
\end{figure}

\begin{figure}
\includegraphics[width=90mm]{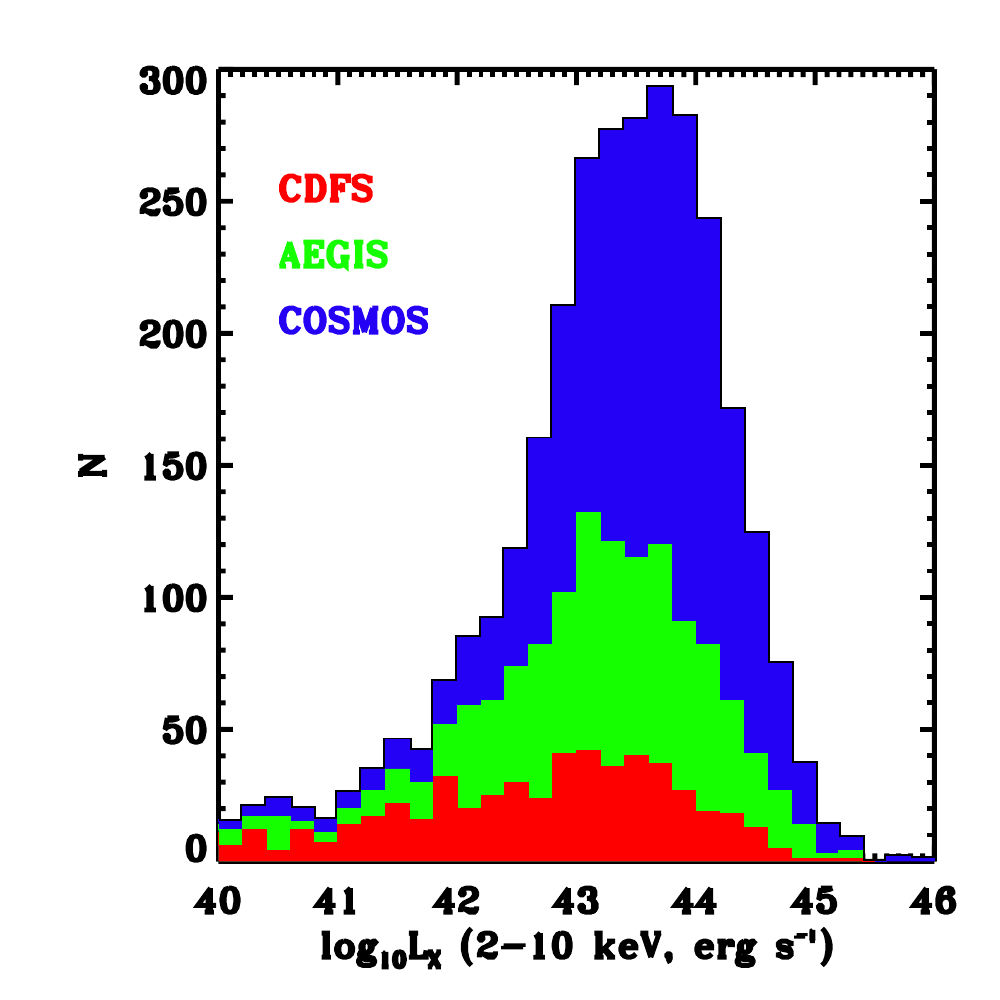}
\caption{Distribution of the intrinsic 2-10 keV X-ray luminosity for all sources in the sample. The logarithmic mean of this distribution is \meanlx. For CDFS this is \meanlxa, for AEGIS it is \meanlxb\ and for COSMOS it is \meanlxc.}
\label{fig_lxdist}
\end{figure}

Finally, one of the most important parameters which can be derived from X-ray spectral fitting is the intrinsic X-ray luminosity. This depends on the flux measured in the X-ray spectrum, the redshift and \nh. Fig. \ref{fig_zlxnh} shows how these samples cover the \lx-$z$ and \lx-\nh\ planes. Combining the relative depths and areas of these surveys gives a wide range of coverage on these planes, however, due the the fact that they are X-ray selected, there are still significant selection effects against low-luminosity high-redshift sources, and low luminosity-high \nh\ sources. Fig. \ref{fig_fxdist} shows the observed flux distribution of the combined sample, whereas Fig. \ref{fig_lxdist} shows the overall distribution of intrinsic X-ray luminosities.

\subsection{Compton thick AGN}

\begin{figure}
\includegraphics[width=90mm]{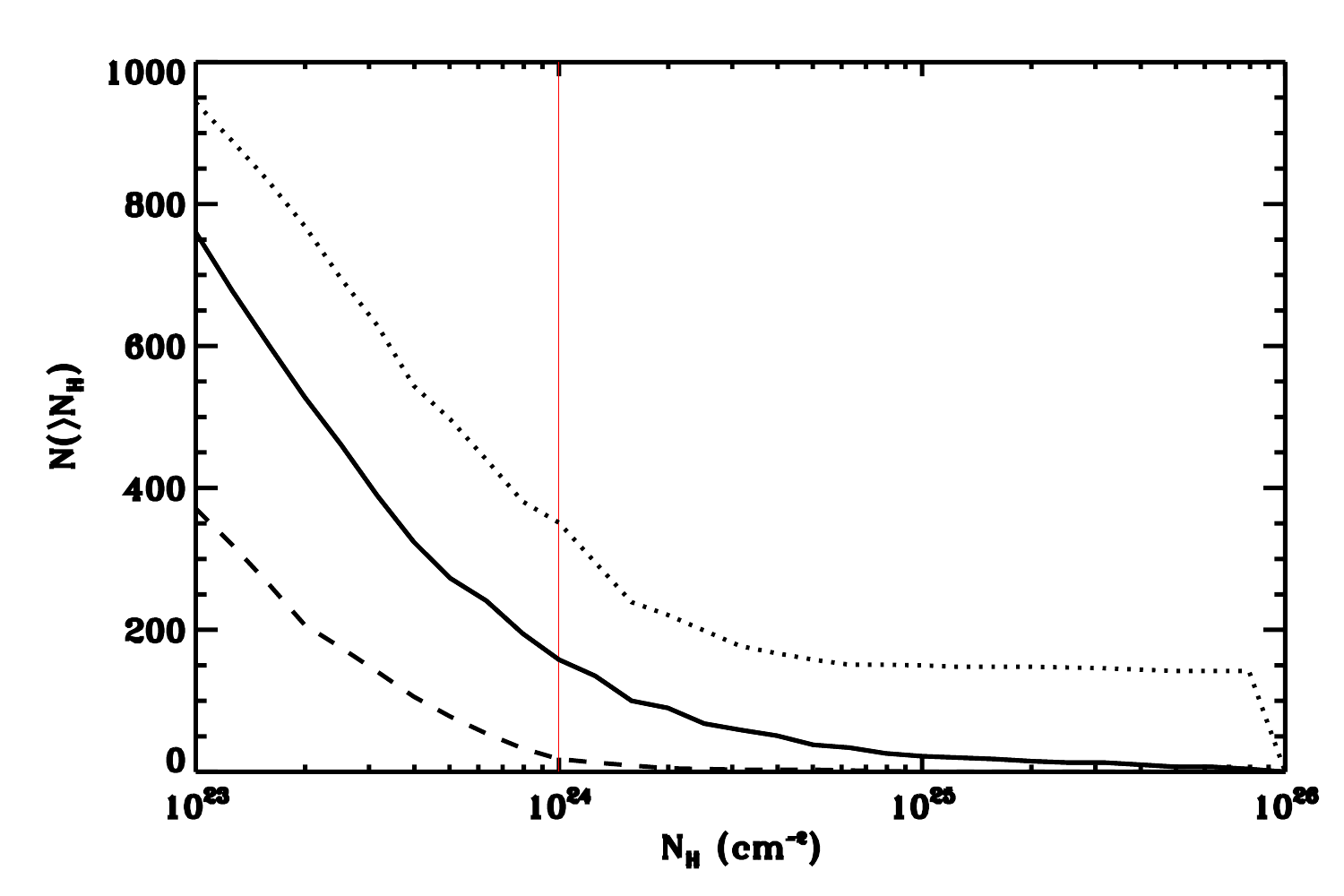}
\caption{The number of sources with best-fit \nh\ greater than a limiting \nh\ (solid line) along with the number of sources with \nh\ constrained above above a limiting \nh\ (90\% confidence, dashed line) and the number of sources where the upper constraint on \nh\ is above a limiting \nh\ (90\% confidence, dotted line). The red line is at \nh=10$^{24}$ \cmsq, where \nctagn\ sources have best-fit \nh$>10^{24}$ \cmsq, \ncctagn\ can be constrained to have \nh$>10^{24}$ \cmsq, and \npctagn\ are consistent with \nh$>10^{24}$ \cmsq.}
\label{fig_nhcon}
\end{figure}

The strict definition of a CT AGN is and AGN with \nh$>1.5\times10^{24}$ \cmsq, which is the \nh\ that the obscuring medium becomes optically thick to Compton scattering. However the photon statistics available to us from the {\it Chandra} data do not allow for such a fine distinction. Indeed, due to the inherent low-count nature of these sources, the constraints on \nh\ are often poor. This is illustrated in Figure \ref{fig_ctsnh}, which shows the 90\% confidence range of the best-fit \nh\ value as a function of spectral counts. Figure \ref{fig_nhcon} shows the number of sources with best fit \nh\ as a function of limiting \nh. This reveals that \nctagn\ sources have best-fit \nh$>10^{24}$ \cmsq. Also shown is the number of sources where the \nh\ can be constrained above a certain value at  90\% confidence, with only \ncctagn\ sources constrained to have \nh$>10^{24}$ \cmsq. However, given the upper limit on the best-fit \nh, as many as \npctagn\ are consistent with \nh$=10^{24}$ \cmsq. This presents a large range in the possible numbers of CT AGN in our sample.

For our stated goals, we thus define a sample of `highly probable' CT AGN having a best fit \nh$>10^{24}$ \cmsq\ and where the \nh\ is constrained to be greater than 10$^{23.5}$ \cmsq. These number \ngctagn\ in total, \ngctagna\ in CDFS, \ngctagnb\ in in AEGIS and \ngctagnc\ in COSMOS. Henceforth, we described these as CT AGN, and describe the remaining sources as those with best-fit \nh$>10^{24}$ \cmsq.

We present a detailed look at a selection of 24 CT AGN and sources with best-fit \nh$>10^{24}$ \cmsq\ newly identified in this work, eight from each of the three surveys utilised. We select these to have high quality redshifts, being spectroscopic, or a photometric redshift where the peak in the probability distribution, $P(z)>90\%$, so that the finer features of the spectra, such as the Fe K$\alpha$ line, can be seen. As already mentioned, a defining feature of the X-ray spectrum of CT AGN is a high equivalent width (EW) Fe K$\alpha$ line, which have been typically measured in excess of 1 keV. The torus models that we use in this work include the Fe K$\alpha$ line self-consistently, and thus cannot be used to calculate the EW. For these 24 sources, we estimate the EW by refitting the spectra with an absorbed power-law ({\tt zwabs*power}), plus a narrow gaussian line component at fixed rest frame 6.4 keV and a secondary power-law component. We estimate the uncertainty on the EW from the uncertainty in the normalisation of the line component. The results of these calculations are presented in Table \ref{tab_CT AGN}, along with the details of the torus fits. We find that the majority (14/24) of these sources do indeed exhibit Fe K$\alpha$ EW$>$ 1 keV, although again these are not well constrained. All spectra with the exception of CDFS 430 are consistent with at least EW=600 eV, which is the minimum EW of the CT sources predicted by Monte-Carlo simulations of a torus geometry \citep{ghisellini94,murphy09,brightman11}. Interestingly enough, CDFS 430 has a well constrained \nh\ ($\Delta$\nh$\simeq0.3$) despite its low Fe K$\alpha$ EW. This could be explained with a low covering factor of the torus. Indeed, this spectrum is best fit by the torus model with 60$^\circ$ opening angle. Unfortunately the photon statistics do not allow us to constrain the opening angle directly.

Included in our results, we find \nhctagn\ sources where the best fit \nh$>10^{25}$ \cmsq, which can be referred to as reflection dominated CT AGN or heavily Compton thick AGN due to transmitted emission being suppressed such that only reflected emission can be seen. The number of these sources with respect to mildly CT sources is highly uncertain even in the local universe. Unfortunately photon statistics here do not allow us to say definitively that these sources are heavily CT, with only 3 of these sources with \nh\ constrained to be $>10^{24.5}$ \cmsq.

Finally, we find 23 sources where we find the best fit model to be a simple power-law, but constrain $\Gamma$ to be less than 1.2. While these could be a class of Blazars, known to have intrinsically flat X-ray spectra, it is more likely that these sources are heavily absorbed, perhaps reflection dominated sources, but not identified here as such with our spectral fitting technique. Upon further investigation, we find that these  spectra are not well fit by our models, possibly due to complex absorption. We test the reflection dominated scenario by fitting a pure reflection model \citep[pexmon, ][]{nandra07}, however this also does not provide an improved fit over a simple power-law. Due to these uncertainties, we do not include these in our sample of CT AGN.

\begin{figure}
\includegraphics[width=90mm]{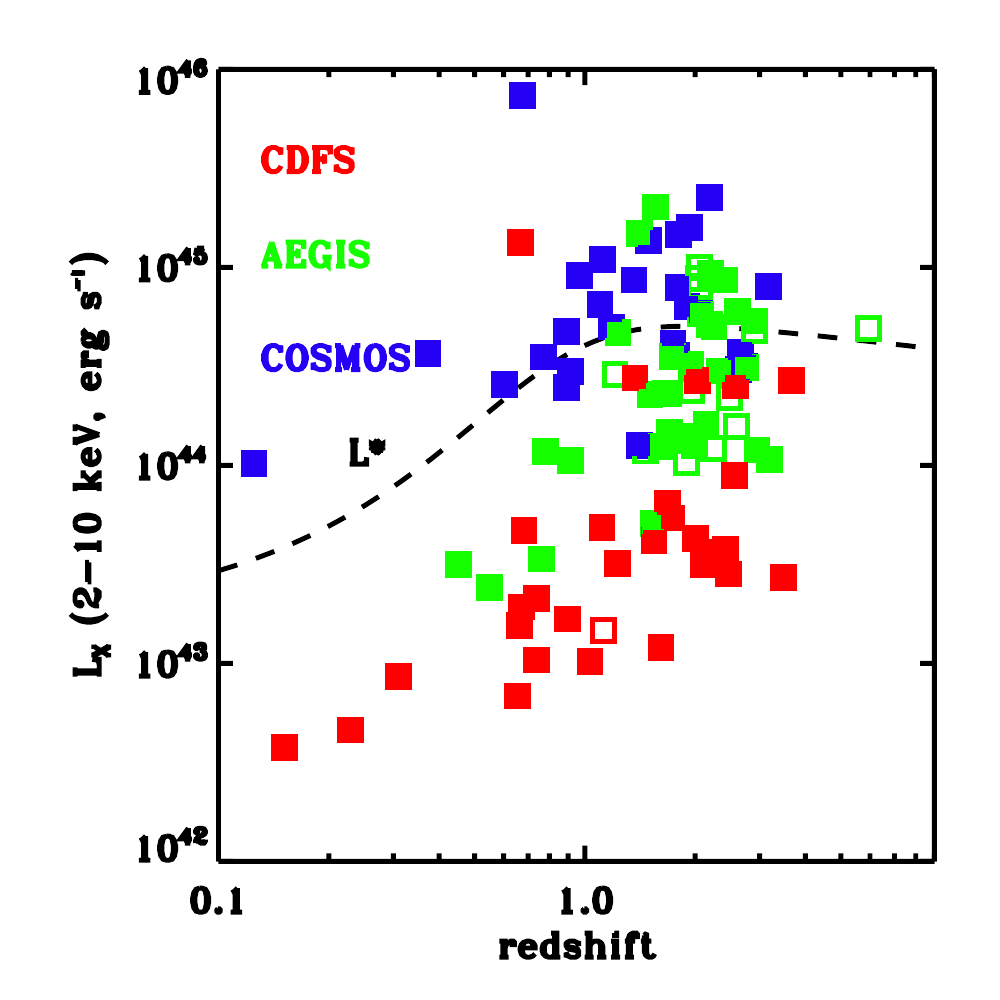}
\caption{Intrinsic rest-frame 2-10 keV \lx-z plot of the \ngctagn\ CT AGN found in this study. Sources with unreliable photometric redshifts are plotted with open squares. The dashed line shows the evolution of $L_{\rm X}^{*}$ from Aird et al (2010).}
\label{fig_ctzlx}
\end{figure}

Fig. \ref{fig_ctzlx} shows the distribution of the CT AGN on the \lx-z plane, and how using the combination of these three surveys results in a large dynamical range in both \lx\ and redshift. We show sources with unreliable photometric redshifts with open squares ($P(z)<$70\%). Our sample of CT AGN spans an intrinsic 2-10 keV luminosity range of $10^{42}\sim3\times10^{45}$ \ergs. The sample also includes \nctagnhiz\ CT AGN at z$>$3.

\begin{figure*}
\includegraphics[width=160mm]{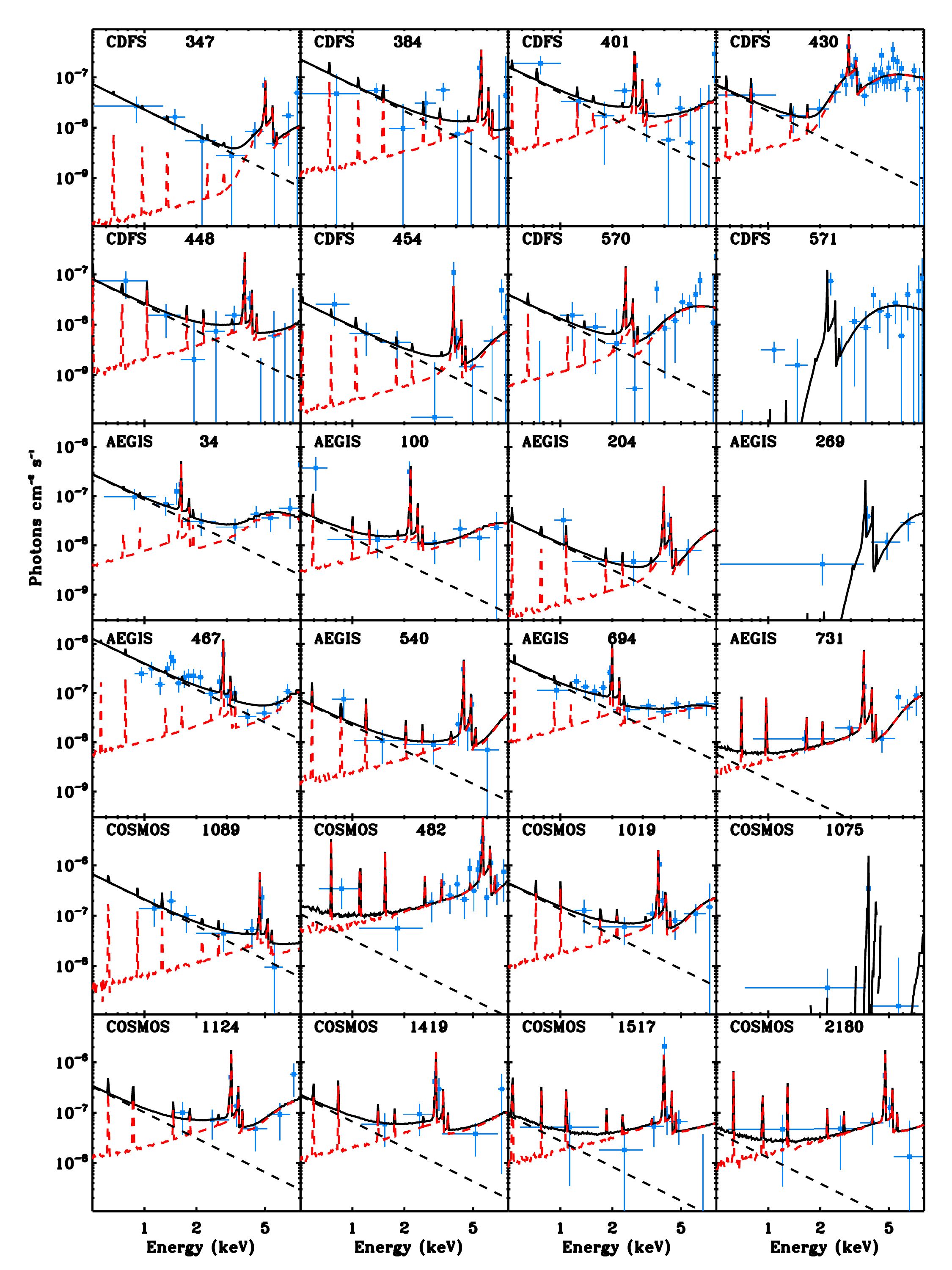}
\caption{Example spectra of 24 CT AGN newly identified in this work, eight from each survey used. Red dashed lines show the torus model which represents the heavily obscured primary emission, whereas dashed black lines show the secondary power-law component, representing intrinsic scattered emission. The spectra have been rebinned for plotting purposes. }
\label{fig_ctspec}
\end{figure*}

\begin{table*}
\centering
\caption{Details of a selection of 24 CTAGN newly identified in this work, eight from each survey used. These are chosen to have high quality redshift information to enable a more detailed look at the EW of the Fe K$\alpha$ line. Column (1) lists the name of the source; column (2) gives the number of spectral counts; column (3) gives the redshift of the source; column (4) gives the peak value of the probability distribution function (\%) of the photometric redshift. 101 indicates a spectroscopic redshift; column (5) gives the best fit log$_{10}$(\nh/\cmsq) value with associated 90\% limits; column (6) gives the EW of the Fe K$\alpha$ line (keV), determined by a fit with a narrow Gaussian;  column (7) gives the observed log$_{10}(F_{\rm X}$/\ergcms, 0.5-8 keV) and column (8) gives the unabsorbed 2-10 keV luminosity.}
\label{tab_ctagn}
\begin{center}
\begin{tabular}{l r r r l l l l}
\hline
Source name & counts & redshift & P(z) & \nh & EW (Fe K$\alpha$) & $F_{\rm X}$ & \lx \\
(1) & (2) & (3) & (4) & (5) & (6) & (7) & (8)\\
\hline
CDFS     347&  247& 0.28&   101& $ 24.0^{+  2.0}_{-  0.6}$ & $ 1.15^{+ 1.29}_{- 1.15}$ & -15.30& $42.06^{+ 0.87}_{- 0.83}$ \\
CDFS     384&  507& 0.15&   101& $ 24.6^{+  1.4}_{-  0.9}$ & $ 3.01^{+ 2.84}_{- 2.22}$ & -14.98& $42.57^{+ 0.36}_{- 1.35}$ \\
CDFS     401&  590& 1.37&   101& $ 26.0^{+  0.0}_{-  2.0}$ & $ 0.41^{+ 0.44}_{- 0.35}$ & -14.84& $44.44^{+ 0.24}_{- 0.43}$ \\
CDFS     430&  847& 1.19&   101& $ 24.0^{+  0.2}_{-  0.3}$ & $ 0.21^{+ 0.15}_{- 0.13}$ & -14.31& $44.26^{+ 0.89}_{- 1.00}$ \\
CDFS     448&  528& 0.68&   101& $ 24.8^{+  1.2}_{-  1.3}$ & $ 0.84^{+ 0.74}_{- 0.60}$ & -15.22& $43.67^{+ 0.26}_{- 1.31}$ \\
CDFS     454&   75& 0.65&   101& $ 24.3^{+  1.7}_{-  0.7}$ & $ 3.35^{+ 3.58}_{- 2.39}$ & -15.61& $42.84^{+ 0.46}_{- 1.25}$ \\
CDFS     570&  437& 1.68&    93& $ 24.3^{+  0.4}_{-  0.5}$ & $<0.86$ & -15.02& $43.81^{+ 0.44}_{- 0.70}$ \\
CDFS     571&  537& 1.90&    92& $ 24.2^{+  0.3}_{-  0.4}$ & $ 0.10^{+ 2.17}_{- 0.10}$ & -15.02& $43.77^{+ 0.44}_{- 0.52}$ \\
AEGIS      34&  119& 2.91&   101& $ 25.0^{+  1.0}_{-  1.4}$ & $ 0.24^{+ 0.36}_{- 0.24}$ & -14.67& $44.73^{+ 0.32}_{- 0.72}$ \\
AEGIS     100&   36& 1.95&    90& $ 25.4^{+  0.6}_{-  1.5}$ & $ 4.30^{+ 4.83}_{- 3.21}$ & -14.94& $44.51^{+ 0.26}_{- 1.39}$ \\
AEGIS     204&   12& 0.61&   101& $ 24.2^{+  1.8}_{-  1.0}$ & $ 0.56^{+ 1.84}_{- 0.56}$ & -15.20& $43.11^{+ 0.66}_{- 1.37}$ \\
AEGIS     269&   18& 0.76&   101& $ 24.2^{+  0.4}_{-  0.6}$ & $ 1.03^{+ 1.81}_{- 1.03}$ & -14.91& $43.53^{+ 1.04}_{- 0.96}$ \\
AEGIS     467&  154& 1.23&   101& $ 24.8^{+  1.2}_{-  0.4}$ & $ 0.62^{+ 0.48}_{- 0.35}$ & -14.30& $44.67^{+ 0.22}_{- 0.42}$ \\
AEGIS     540&   28& 0.45&   101& $ 24.4^{+  1.6}_{-  0.5}$ & $ 1.38^{+ 1.38}_{- 0.94}$ & -14.92& $43.50^{+ 0.33}_{- 0.90}$ \\
AEGIS     694&   91& 2.20&    99& $ 24.9^{+  1.1}_{-  1.0}$ & $ 0.24^{+ 0.37}_{- 0.24}$ & -14.51& $44.97^{+ 0.22}_{- 1.30}$ \\
AEGIS     731&   32& 0.78&   101& $ 24.3^{+  0.2}_{-  0.4}$ & $ 1.45^{+ 4.53}_{- 0.87}$ & -14.62& $44.07^{+ 0.23}_{- 0.69}$ \\
COSMOS    1089&   24& 0.37&   101& $ 25.7^{+  0.3}_{-  3.3}$ & $ 1.90^{+ 2.76}_{- 1.69}$ & -14.61& $43.67^{+ 0.39}_{- 2.17}$ \\
COSMOS     482&   58& 0.12&   101& $ 25.3^{+  0.7}_{-  1.2}$ & $ 1.20^{+ 0.82}_{- 0.63}$ & -13.60& $44.01^{+ 0.12}_{- 0.47}$ \\
COSMOS    1019&   42& 0.73&   101& $ 24.0^{+  2.0}_{-  0.6}$ & $ 0.70^{+ 0.84}_{- 0.55}$ & -14.15& $43.93^{+ 1.00}_{- 0.72}$ \\
COSMOS    1075&   21& 0.68&   101& $ 25.1^{+  0.9}_{-  1.3}$ & $ 4.37^{+ 6.29}_{- 3.86}$ & -15.08& $45.87^{+ 1.21}_{- 1.39}$ \\
COSMOS    1124&   49& 1.02&   101& $ 24.5^{+  1.5}_{-  1.0}$ & $ 1.32^{+ 1.61}_{- 1.14}$ & -14.14& $44.75^{+ 0.36}_{- 1.36}$ \\
COSMOS    1419&   30& 1.10&     100& $ 24.8^{+  1.2}_{-  0.9}$ & $ 5.14^{+ 5.11}_{- 3.24}$ & -14.36& $44.81^{+ 0.30}_{- 1.03}$ \\
COSMOS    1517&   61& 0.60&   101& $ 24.9^{+  1.1}_{-  1.3}$ & $ 1.67^{+ 1.42}_{- 1.02}$ & -14.49& $44.41^{+ 0.24}_{- 1.20}$ \\
COSMOS    2180&   51& 0.35&   101& $ 24.7^{+  1.3}_{-  1.3}$ & $ 1.07^{+ 1.47}_{- 0.97}$ & -14.45& $44.00^{+ 0.25}_{- 1.45}$ \\
\hline
\end{tabular}
\end{center}
\end{table*}

 We present some basic data concerning the identification of \nh$>10^{24}$ \cmsq\ sources per survey, presented in Table \ref{cteff_table}. As the three survey strategies are somewhat different, ranging from the ultra deep and narrow CDFS, to the medium deep and medium wide AEGIS, to the wide and shallow COSMOS field, it is useful to compare what these surveys yield in terms of \nh$>10^{24}$ \cmsq\ sources, especially for future survey planning. We find \nctagn\ of these out of a parent sample of \nsrc\ sources, giving an observed fraction of \ctfrac\%. Individually, for CDFS, AEGIS and COSMOS this is \ctfraca, \ctfracb\ and \ctfracc\% respectively. We stress that these are the observed fractions of sources with \nh$>10^{24}$ \cmsq\ identified. In order to calculate the intrinsic fractions, survey biases and sensitivity, along with spectral fitting effectiveness must be taking into account. This was done in BU12 for the CDFS and compared to X-ray background synthesis models. We do not make a comparison to XRB models here as we do not focus on the intrinsic CT fraction, rather on compiling a robust sample of CT AGN.
  
In terms of the detection and identification efficiency of sources with \nh$>10^{24}$ \cmsq, in units of number detected and identified, per unit exposure time per unit area (Ms$^{-1}$ degree$^{-2}$), CDFS, AEGIS and COSMOS yield \cteffa\, \cteffb\ and \cteffc\ sources Ms$^{-1}$ degree$^{-1}$. This indicates that medium depth surveys at $\sim$Ms depths, such as AEGIS-XD, are most efficient at detecting and identifying sources with \nh$>10^{24}$ \cmsq.

\begin{table*}
\centering
\caption{Summary and analysis of the Compton thick population by survey. Column (1) gives the survey name, column (2) gives the total survey exposure time in Ms, column (3) gives the area of the survey in armin$^{2}$, column (4) gives the sensitivity of the survey in the 0.5-8 keV band, in units of \ergcms, column (5) gives the number of all sources with \lx$>10^{42}$ \ergs, column (6) gives the number of sources with best-fit \nh$>10^{24}$ \cmsq\ and \lx$>10^{42}$ \ergs, column (7) gives the number of sources with best-fit \nh$>10^{25}$ \cmsq\ and \lx$>10^{42}$ \ergs, column (8) gives the fraction of all sources with best-fit \nh$>10^{24}$ \cmsq, column (9) gives the efficiency for each survey in units of the number of sources with \nh$>10^{24}$ \cmsq\ detected and identified, per unit exposure time per unit area (Ms$^{-1}$ degree$^{-1}$) and column (10) gives the mean intrinsic 2-10 keV luminosity of the sources with best-fit \nh$>10^{24}$ \cmsq\ in each sample (log \ergs)}
\label{cteff_table}
\begin{center}
\begin{tabular}{l r r r r r r c c c}
\hline
Survey & Exposure & Area & Sensitivity & \nh$\geq 10^{20}$ & \nh$\geq10^{24}$ & \nh$\geq10^{25}$ & $\frac{N_{\rm H}\geq 10^{24}}{N_{\rm H}\geq 10^{20}}$ & efficiency & $<$log$_{10}($\lx$)/$\ergs$>$ \\
(1) & (2) & (3) & (4) & (5) & (6) & (7) & (8) & (9) & (10) \\
\hline
CDFS    & 4   & 465   & $3.2\times10^{-17}$     &  377 &   36 &    4 &    9.5\% &   69 &   43.6\\
AEGIS     & 2.4 & 871   & $1.7\times10^{-16}$   &  809 &   57 &    9 &    7.0\% &   98 &   44.3\\
COSMOS    & 1.8 & 3240  & $5.7\times10^{-16}$   & 1553 &   64 &    9 &    4.1\% &   39 &   44.7\\
Combined  & 8.2 & 4576  & -                     & 2739 &  157 &   22 &    5.7\% &   15 &   44.3\\
\hline
\end{tabular}
\end{center}
\end{table*}

\subsection{Nature of the obscuration}
\label{sec_nature}

The nature of the obscuring material in AGN beyond the local universe is a matter of much debate. One particular topic concerns the question of to what extent AGN obscuration is linked to the properties of the host galaxy. Furthermore, the fraction of obscured sources have been found to be dependent on the luminosity of the AGN, and that this fraction evolves with redshift, suggesting, for example, an evolution of the putative torus properties. Our main tools in this study are the new torus models which we employ in our spectral fitting which can give us information about the covering factor of the obscuring material. As these spectral models are specifically designed to account for the geometry of the obscuring material, we can make inferences on these of the evolution of the covering factor with redshift, and its dependence on luminosity. 

While in our analysis, we do not constrain the torus opening angle directly, as the spectral quality is not high enough, we do test three different fixed covering factor scenarios, specifically torus opening angles of 0\degree, 30\degree\ and 60\degree, when fitting the spectra. Fig \ref{fig_mospec} gives an example of each of these. By investigating how the best fit model from these three scenarios changes with luminosity and redshift, we can further investigate the relationship between the nature of the obscuring material and these parameters. 

For this investigation, when considering spectral fits with these torus models we consider only AGN which we have found to be heavily obscured, with \nh$>10^{23}$ \cmsq. It is at these column densities where the optical depth to Compton scattering becomes significant, and where the torus models are well suited.  In the previous section of this paper we focussed on the robust identification of heavily obscured AGN, taking care as to minimise the contamination by unabsorbed sources. However, in the following section we aim to characterise these heavily obscured sources with respect to source luminosity and redshift on a statistical basis and therefore require as large a sample as possible. For this reason we lower the significance threshold in the spectral fit for the addition of free parameters from $\Delta$c-stat of 2.71 to 1.0. In doing so we increase the sample of heavily obscured sources from \nhabs\ to \nhabstor.

\begin{figure}
\includegraphics[width=90mm]{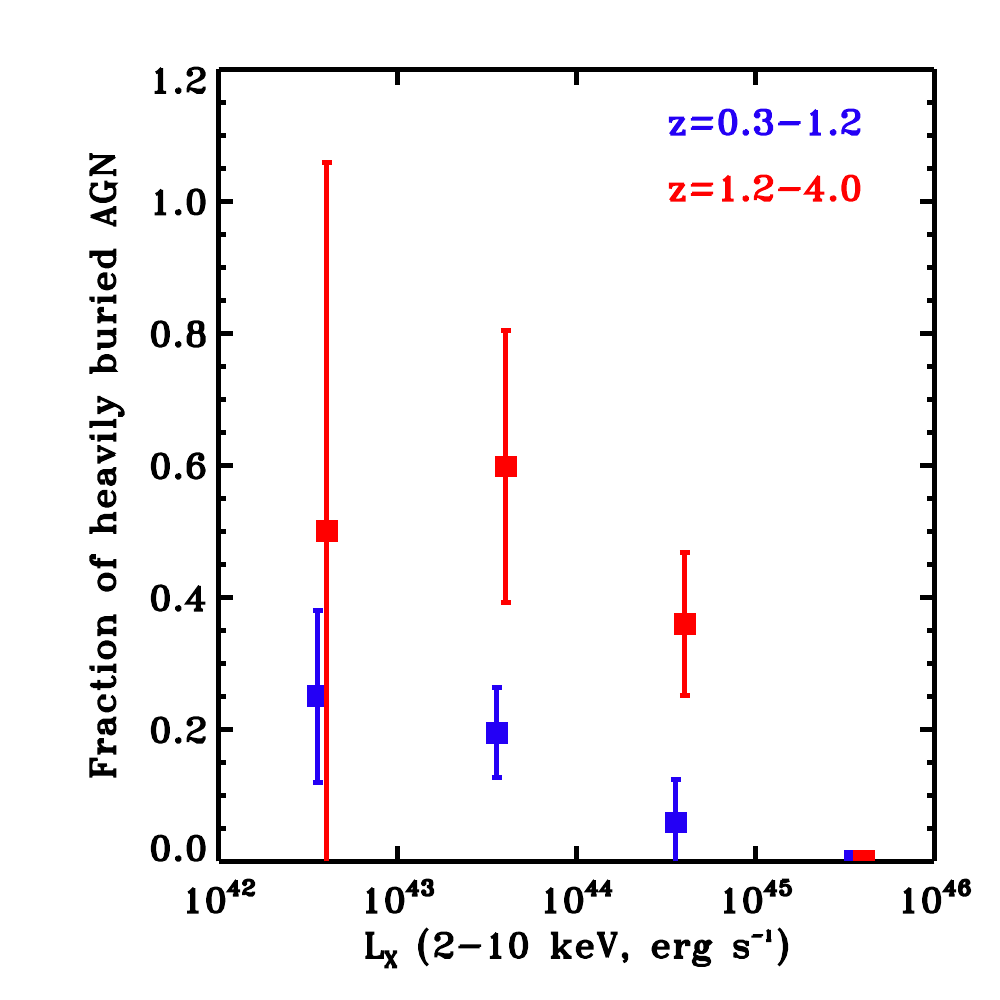}
\caption{The fraction of heavily buried AGN as a function of intrinsic rest-frame 2-10 keV luminosity for two redshift bins. The fraction is defined as the fraction of spectra best fit by the torus model with a zero degree opening angle with respect to the torus model with a 30 or 60 degree opening angle. Errors on the fraction are poisson errors.}
\label{fig_ftor}
\end{figure}

In Fig. \ref{fig_ftor}, we plot the fraction of sources which are best fit by the torus model with a zero degree opening angle, with respect to those best fit by the torus model with opening angles of 30 or 60 degrees as a function of intrinsic X-ray luminosity and for two redshift bins. The zero degree opening angle spectra represent a scenario where the source is heavily buried in material which is further supported by the absence of any significant soft excess component. We therefore denote this fraction as the `fraction of heavily buried AGN'. These heavily buried AGN show no evidence for soft scattered emission or for a Compton reflected component originating from the wall of the torus. What is shown is a decreasing fraction of these sources towards higher luminosities. A possible interpretation of this is that the covering factor of the obscuring material decreases with source luminosity, in agreement with the receding torus model. What is also shown is that this fraction increases with redshift at a common luminosity, consistent with the evolution of the obscured fraction with redshift. We point out that the average fraction of heavily buried AGN presented in this figure is lower than that presented in Table \ref{tab_models} due to the lower $\Delta$c-stat criterion used for the figure, resulting in more spectra best fit by model A or B.

\begin{figure}
\includegraphics[width=90mm]{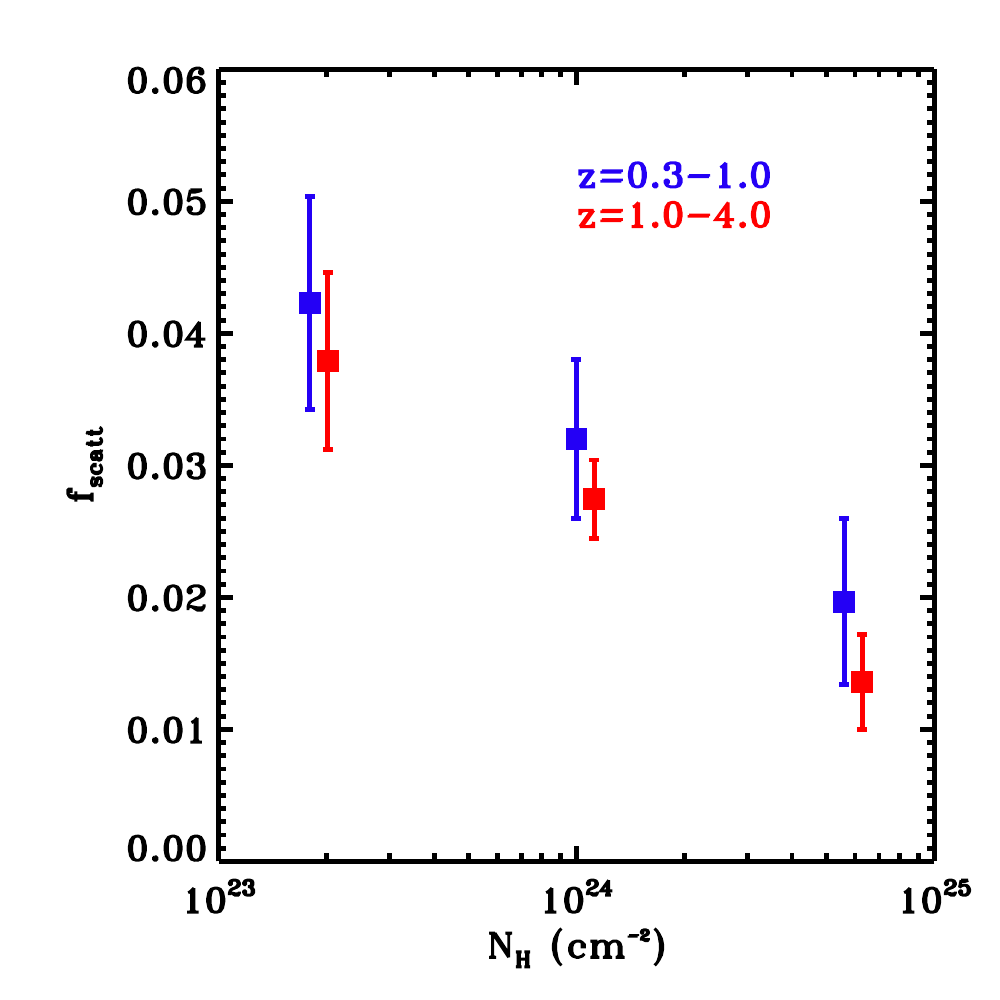}
\caption{The mean scattered fraction, \fscatt, as a function of \nh. This analysis is done in the rest frame 2-10 keV band, however we restrict the analysis to where \fscatt$>0$ in order to avoid the bias against low \fscatt\ where \nh\ is low and hence the scattered emission is not visible. The errors are standard errors on the mean. }
\label{fig_fscatt2}
\end{figure}

Fig \ref{fig_fscatt2} plots the mean \fscatt\ as a function of \nh. It was shown in BU12 that as \nh\ increases, \fscatt\ decreases, implying that the most heavily absorbed sources are also the most heavily buried. We confirm this trend here in two redshift bins.

\begin{figure}
\includegraphics[width=90mm]{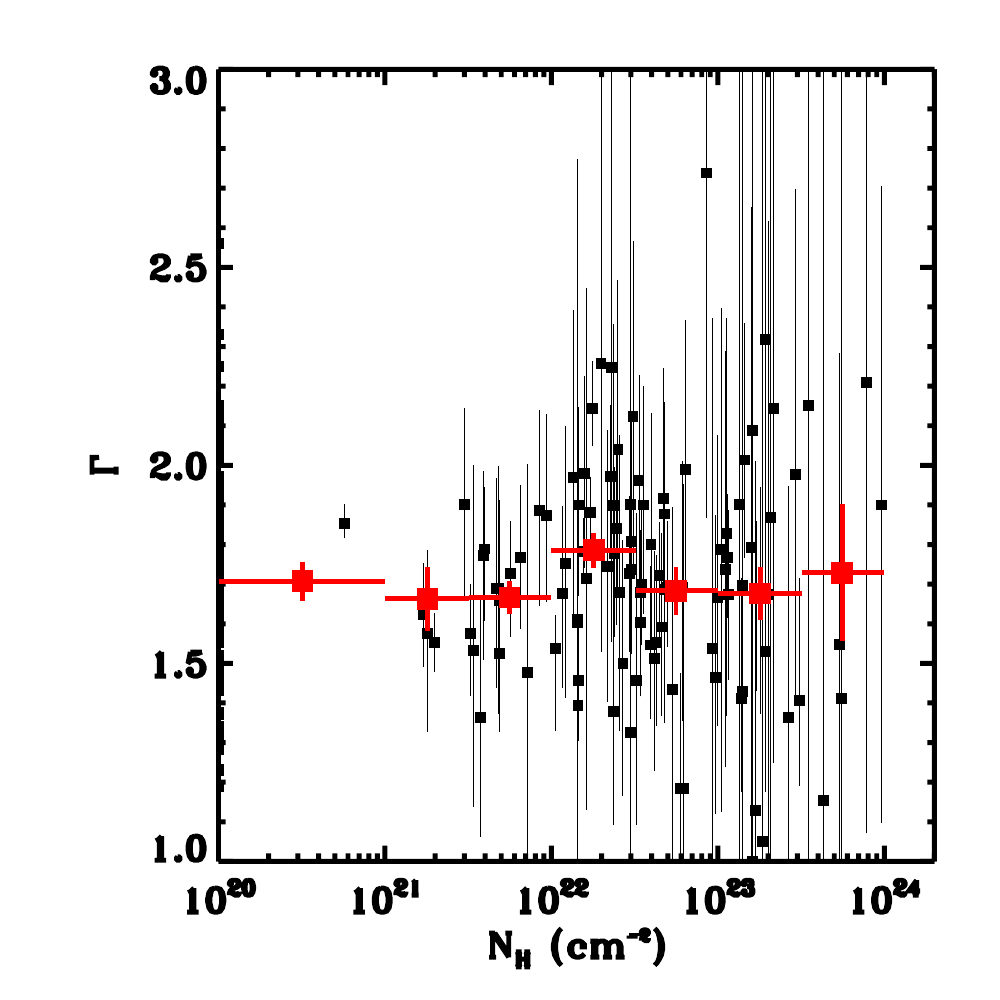}
\caption{$\Gamma$ vs. \nh, where the black crosses and error bars show individual measurements for sources greater than 600 counts. Red data points are binned averages and standard errors. The first bin includes sources where \nh=0.}
\label{fig_gamnh}
\end{figure}

It has been shown that the photon index, $\Gamma$, measured in the X-ray spectrum of AGN is correlated well with the Eddington ratio, \lamedd\ \citep{shemmer08,risaliti09b,brightman13}. We use this relationship to assess if we can infer any relationship between the parameters of obscuration and the Eddington ratio. In Table \ref{tab_models} we presented the mean $\Gamma$ for the torus models with three different covering factors. We see that as the covering factor of the torus increases, the average $\Gamma$ also increases, from $<\Gamma>=$\gamtora\ for spectra with a 60\degree\ opening angle to $<\Gamma>=$\gamtorb\ for spectra with a 30\degree\ opening angle and $\Gamma>=$\gamtorc\ for spectra with a 0\degree. A possible interpretation of this is that the most buried AGN are accreting with higher Eddington ratios. We also investigate if there is any trend between $\Gamma$ with \nh. We plot this data in Fig. \ref{fig_gamnh}, where we also show the binned averages of $\Gamma$ for several \nh\ bins. This shows a constant relationship between $\Gamma$ and \nh, which may imply that there is no dependance of Eddington ratio on the line of sight obscuration.

\section{Discussion}
\label{sec_discuss}

\subsection{Spectral fitting method}
\label{discussion_specfit}

In our spectral fitting method, we use techniques where their reliability and statistical robustness have not to date been clearly determined. These are the use of the Cash statistic with background subtracted spectra, the random grouping of the spectra, the $\Delta$c-stat criterion for model selection and the use of torus models. We investigate and discuss how the use of these may affect the identification of CT AGN using simulations and comparisons to other work.

The use of the Cash statistic is not strictly suitable for spectra where background subtraction has taken place, however {\sc xspec} has a modified version of the Cash statistic which is appropriate for spectra in the presence of a background, which we use. We compare our spectral fit method with two other works which employ two different spectral fitting methods in order to assess if any biases have taken place. \cite{rangel14} (R14) have presented results from X-ray spectral fitting of massive galaxies at $z\sim2$ in the CDFS. They also utilise the torus models of BN11, and carry out fitting with the Cash statistic in the presence of a background, however they utilise fixed energy binning, rather than the minimum of one count per bin that we use. This has the advantage of being unbiased towards the count distribution in the spectrum. We compare our results on the measurement of \nh\ for five sources which R14 has identified as CT in Table \ref{specmeth_table}. Our results are in excellent agreement. Furthermore, B14 have employed an entirely different spectral fitting method, again using the torus models of BN11, but based on Bayesian techniques in which they model the background separately and thus do not group the spectrum at all. Again, these results are in very good agreement with both our results and R14. 

\begin{table}
\centering
\caption{Comparison of the measurement of \nh\ for five sources identified as CT in R14 for different spectral fitting methods. Column (1) gives the source name, column (2) gives the \nh\ with uncertainties from this work, while columns (3) and (4) give the \nh\ presented in R14 and B14 respectively. }
\label{specmeth_table}
\begin{center}
\begin{tabular}{l l l l}
\hline
Source & this work & R14 & B14 \\
(1) & (2) & (3) & (4) \\
\hline
cdfs4Ms\_226 & $24.14^{+ 0.29}_{- 0.26}$ & 24.18 & 24.2 \\
cdfs4Ms\_271 & $24.32^{+ 0.38}_{- 0.74}$ & 24.50 & 24.4 \\
cdfs4Ms\_437 & $24.12^{+ 0.25}_{- 0.30}$ & 24.16 & 24.3 \\ 
cdfs4Ms\_460 & $24.11^{+ 0.35}_{- 0.44}$ & 24.17 & 24.4 \\
cdfs4Ms\_474 & $24.03^{+ 0.37}_{- 0.44}$ & 24.33 & 24.2 \\
\hline
\end{tabular}
\end{center}
\end{table}

Finally, we conduct simulations to investigate if randomly grouping the spectra, as we have done, has any systematic effect on our results. We compare this to fitting the ungrouped spectrum. We simulate 100 spectra of aegis\_540, a CT AGN with \nh=2.2$\times10^{24}$ \cmsq, $\Gamma$=1.7 and \fscatt=0.2\%, in the regime of 10-30 counts. We find that for the grouped spectra, where the correct model has been identified, the average \nh=4.0$\times10^{24}$ \cmsq, whereas for the ungrouped spectra, this is 8.5$\times10^{24}$ \cmsq. We infer from this that the grouping does not introduce any systematic bias in our measurement of \nh\ and the identification of CT AGN.

Previous X-ray spectral analysis of sources in our survey fields have been carried out with the 1 Ms data in the CDFS by \cite{tozzi06} and most recently with the 4 Ms data by \cite{buchner14} (B14). To date no spectral analysis has been done in the AEGIS field. \cite{mainieri07} presented a spectral analysis of XMM-COSMOS sources and most recently, \cite{lanzuisi13} (L13) presented a spectral analysis of the brightest ($>$70 net counts) 390 sources in $Chandra$-COSMOS. The analyses of L13 and B14 are the only ones that use the same data sets as our own, including the redshift catalogues utilised, and as such we compare our results with theirs. 

\begin{figure}
\includegraphics[width=90mm]{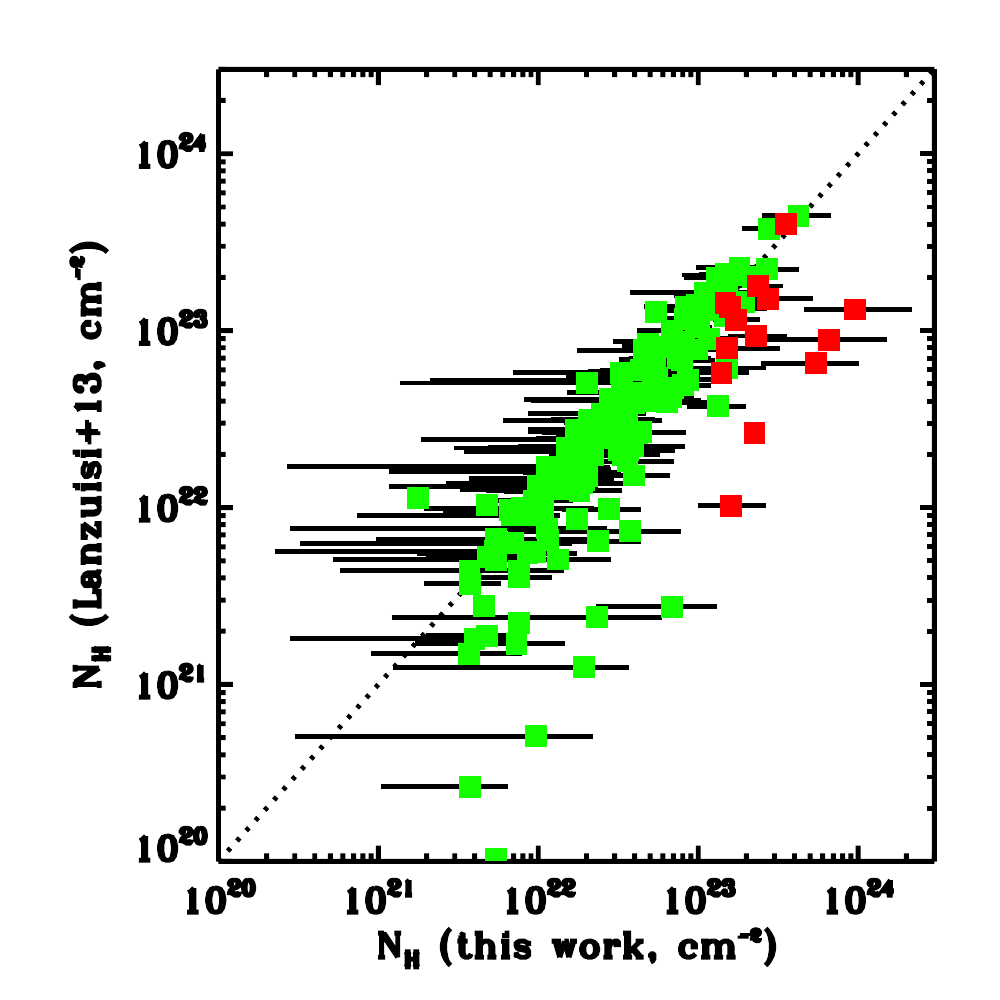}
\caption{Comparison of \nh\ determined in this work to that presented in L13 for 390 of the brightest C-COSMOS sources. Red data points are sources best fit with the torus model plus a scattered component (model A or B) and green data points are sources best fit the torus model without a scattered component (model C).}
\label{fig_lanznh}
\end{figure}

The major difference between our analysis and that of L13 are the spectral models used. L13 use a simple power-law model attenuated by photoelectric absorption, whereas we employ the X-ray spectral torus models of \cite{brightman11}, and where it is required in the fit, a soft scattered component. Therefore in the comparison, we check how the two \nh\ measurements agree. Fig \ref{fig_lanznh} shows the two \nh\ values plotted against each other, with very good agreement within the measurement errors. Red points on this plot show where the torus model with the scattered component has been included in our fit. This produces a systematically higher \nh\ than found by L13, who do not include a soft scattered component. This component accounts for soft X-ray emission that in its absence, would be attributable to lower photoelectric absorption when fit with a simple absorbed power-law.

\begin{figure}
\includegraphics[width=90mm]{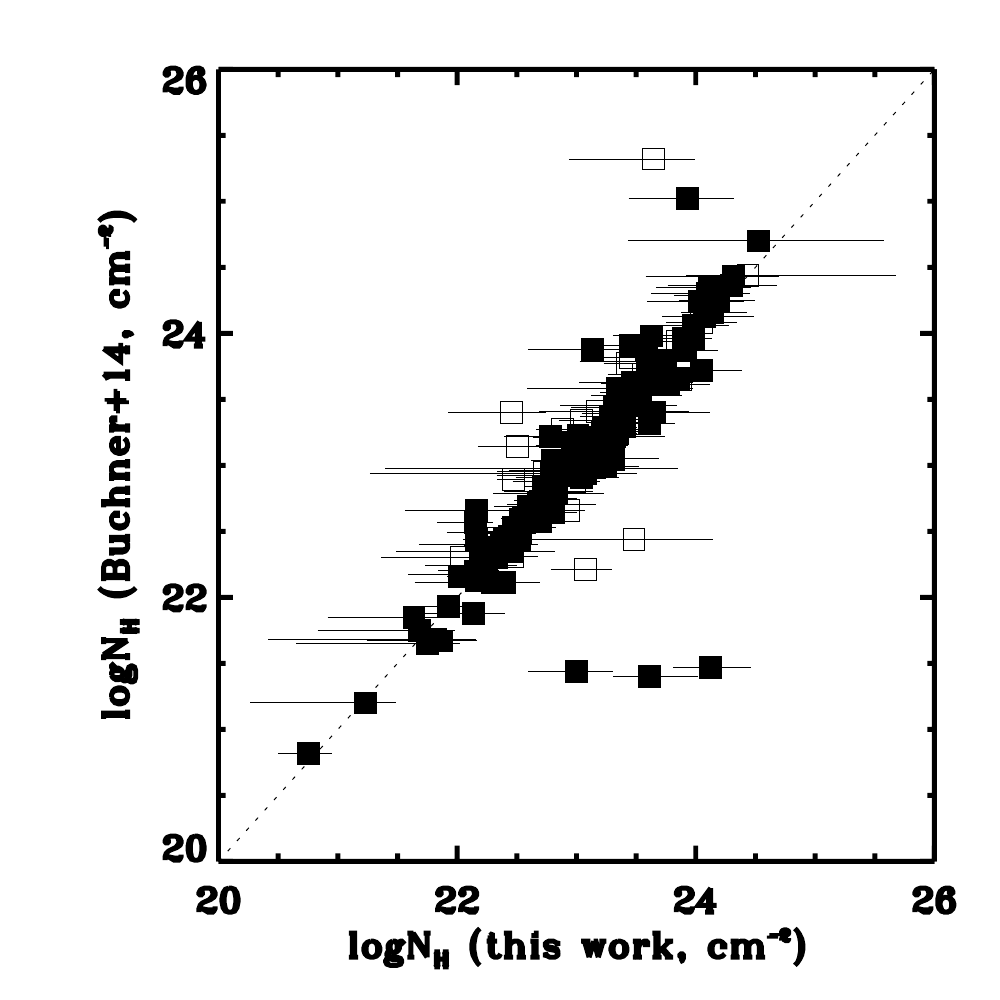}
\caption{Comparison of \nh\ determined in this work to that presented in B14 for sources in the CDFS for sources where \nh\ has been constrained to be greater than zero in our work. Boxes show sources where the peak of the redshift probability function is less than 90\%.}
\label{fig_buchnh}
\end{figure}

The differences between our analysis and that of B14 is that they use a Bayesian based method for model selection and spectral fitting, also modelling the background separately and taking into account the probability distribution in the photometric redshifts. Their Bayesian approach allows them to put priors on the parameters, such as for $\Gamma$ where they use a Gaussian prior peaked at 1.95 with a standard deviation of 0.15, whereas our approach is to fix $\Gamma$ to a single value for low counts, or allow it to take a wide range in values. Their Bayesian technique is statistically more robust than our own for parameter estimation, but as we show, the results are not dissimilar.  Fig. \ref{fig_buchnh} shows the comparison between the two \nh\ measurements. For clarity and fair comparison, we show only sources where the \nh\ could be constrained in our fit. Again here the two \nh\ measurements agree very well. A few outliers result either where the probability distribution function of the photometric redshift is broad, or differing approaches to the determination of $\Gamma$ as discussed produce differing results.

We have assessed the use of our $\Delta$c-stat criterion for model selection, which we described in Section \ref{sec_modsel}. For this we simulated spectra of the four different models used, and fitted these in the same way as the real spectra, concluding that the $\Delta$c-stat criterion used was consistent with the 90\% confidence level in adding additional parameters.

We have also investigated the use of the BN12 colour-colour scheme for identifying CT AGN from hardness ratios. We find that 74\% of the \nh$>10^{24}$ \cmsq\ sources identified here lie within the selection area, while 70\% of the CT AGN are found there. We argue that hardness ratios of this type can be useful for finding CT AGN, however a spectrum will always contain more data than a hardness ratio, and thus should be used over a hardness ratio where possible.

Lastly, we find that the average photon index, $\Gamma$ in our analysis is \meangamml\ with a standard deviation of \sdgamml, which is somewhat flatter than previous works have found. We also find that this does not depend on \nh, even for heavily obscured sources, which indicates that this is not due to incorrectly modelling the obscuration. \cite{buchner14} find in their analysis of CDFS sources using the BN11 torus models, that an additional reflection component, possibly originating from the accretion disc. We therefore surmise that our flat average $\Gamma$ is not intrinsic and due to unmodelled reflection that does not originate in the torus.

\subsection{Compton thick AGN}

The first CT AGN identified at cosmological distances was CDFS-202 \citep{norman02} at a redshift of 3.7, using the 1 Ms \chandra\ exposure of the CDFS. This source corresponds to cdfs4Ms\_047 in our catalogue. In our analysis, we find that \nh=8.2$^{+2.3}_{-1.8}\times10^{23}$ \cmsq, which is in good agreement with this initial measurement. Some of the most recent searches for CT AGN in CDFS have been performed using the deep \xmm\ data also available in that field \citep{comastri11,iwasawa12,georgantopoulos13}. Most recently, \cite{georgantopoulos13} presented a sample of four secure CT AGN from their analysis of these data. These sources are 47, 50, 338, 430 in our catalogue with XMM-IDs of 144, 147, 324, 66 respectively. From our spectral fits of these sources, we find log$_{10}$\nh/\cmsq=23.9, 23.7, 23.9 and 24.0, thus we confirm the CT nature of one of their sources, while in rough agreement with the others. 

\cite{feruglio11} (F11) identified two Compton thick quasars hosted by massive, star-forming {\it BzK} galaxies using the {\it Chandra} 4Ms data from their high EW Fe K$\alpha$ lines. These sources, {\it BzK} 4892 and {\it BzK} 8608 correspond to 77 and 283 in our CDFS catalogue. For source 77, which has a spectroscopic redshift of 2.578, we find that log$_{10}$\nh=$25.7^{+0.3}_{-0.9}$ \cmsq\ with an intrinsic log$_{10}$\lx=44.4 \ergs, which confirms the reflection dominated, quasar nature of this source from F11. For source 283, the photometric redshift of 2.56 that we use leads to a measured log$_{10}$\nh=$23.5^{+0.2}_{-0.3}$ \cmsq, however the spectrum requires $\Gamma=1.0$. F11 use a photometric redshift of 2.9, which when used in our spectral fit gives log$_{10}$\nh=$23.9^{+0.1}_{-0.2}$ \cmsq\ with $\Gamma$ fixed at 1.7, and is thus consistent with the Compton thick nature of this source reported by F11.

In BU12 a spectral analysis of sources detected in the \cite{luo08} 2 Ms catalogue was carried out and a sample of Compton thick AGN was presented. We have extended this analysis to sources detected in the 4 Ms exposure, and in doing so, not only are fainter sources included, but improvements are made in the source positions (due to greater photon statistics), and in the photometric redshifts, as new CANDELS data in the region have been utilised (Hsu et al. 2014 submitted). We identify a further 27 sources as Compton thick with these new data over those presented as secure CT AGN in BU12. Furthermore, there are five sources which we find not to be Compton thick from BU12. Two of these (28 and 221 or 306 and 186 from L08) have new redshifts here, which is the reason we no longer find them to be CT. Source 28 has a new redshift of 1.71 (7.62 previously), where we now find log$_{10}$\nh/\cmsq=23.4. Source 221 has a new redshift of 1.53 (3.98 previously) and we now find log$_{10}$\nh/\cmsq=22.5. Source 266 (145 in L08) had a high scattered fraction in BU12 (44\%). We exclude such sources from being CT here as they are most likely contaminating sources. For sources 126 and 453 (282 and 156 in L08) in BU12, the heavily obscured component was fitted to a hard excess in the spectrum, with $\Gamma$ fixed at 1.9, however here we fix $\Gamma$ to 1.7, which is mean of the $\Gamma$ distribution we find in this work. In doing so, the hard excess is partly accounted for in the harder spectral shape and the addition of a heavily obscured component is not as significant. Thus these sources are fitted with unobscured power-laws here. The above comparisons highlight the sensitivity of CT AGN identification on redshift information. For the most reliable CT AGN searches, robust redshifts are required.

The most recent work to identify CT AGN using the 4Ms {\it Chandra} data in the CDFS is the work by B14. As mentioned above, this work uses a Bayesian based method for model selection and spectral fitting, and the authors model the background separately and take into account the probability distribution in the photometric redshifts. They also use the same torus models that we do, thus a comparison is important. The sample of B14 consists of $\sim350$ AGN in the CDFS, selected from the same catalogue as we use here, however excluding sources with \lx$<3\times10^{42}$ \ergs, effective $\Gamma>1$ and \lx/$L_{\rm opt}<-1$, determined to be dominated by the host galaxy. B14 find that 26 sources have \nh$>10^{24}$, 20 of which we confirm as having \nh$>10^{24}$ \cmsq. Of the six sources where we do not agree with this classification, four of them have a measured \nh\ which is consistent with $10^{24}$ \cmsq\  considering the uncertainties, while one has a flat gamma and no measured \nh, and thus suggestive of heavy absorption. The final source has a photometric redshift with a broad probability distribution. While we only use the peak value, B14 take into account the full distribution, which is likely the cause of the disagreement. Finally we also find six source with \nh$>10^{24}$ \cmsq\ that they do not classify as such. For one of these, again they are consistent within the uncertainties, whereas the remaining five are fairly low significance CT AGN, and would be excluded in our work were a $\Delta$c-stat criterion of 4.0 used, rather than 2.7. We thus conclude that the difference in identifying sources with \nh$>10^{24}$ \cmsq\ between our work and that of B14 is not large considering the statistical uncertainties.

The highest redshift CT AGN known to date was presented by \cite{gilli11}, hosted by an ultra-luminous infrared galaxy at z=4.75. While this source remains formally undetected in the catalogue that we use due to a high level of background, we extract the spectrum at the position given by X11 (XID403). Our spectral analysis of this source confirms its CT nature, with \nh=1.6$\times10^{24}$ \cmsq, and an intrinsic luminosity of 1.27$\times10^{44}$ \ergs. 

While the throughput of \xmm\ has the advantage of obtaining higher signal to noise data over \chandra, especially above 5 keV, its limitation in identifying CT AGN in deep surveys is its sensitivity. The 3.45 Ms \xmm\ exposure in the CDFS yields sources down to a 2-10 keV flux limit of $6.6\times10^{-16}$ \ergcms\ once the data are screened for flares \citep{ranalli13}. The large point spread function (PSF) of \xmm\ does not allow it to probe fainter fluxes due to source confusion. We find that 30/44 (68\%) of these lie below the XMM-CDFS flux limit (where the 2-10 keV flux limit has been converted to 0.5-8 keV using a power-law spectral model with $\Gamma$=1.4). We surmise that the work done with \xmm\ has produced relatively small samples of CT AGN with respect to \chandra\ due to the lack of depth. The next generation X-ray telescope, {\it ATHENA} will achieve at least a 5'' PSF on axis allowing it to go to the depths required to detect and characterise large numbers of CT AGN \citep{nandra13}. The three survey fields considered in this work are also the focus of the {\it NuSTAR} extragalactic survey program. With its sensitivity above 10 keV, {\it NuSTAR} data in these fields will provide further constraints on distant CT AGN \citep[e.g. ][]{alexander13, delmoro14}. 

A detailed determination of the Compton thick fraction at several epochs was presented in BU12, using the CDFS alone and correcting for survey biases, incompleteness and contamination from simulations. This analysis would benefit from the much enlarged sample here, however such analysis is out of the scope of this paper, where we instead focus on the compilation of a clean and robust CT AGN sample such that each source may be taken on a case by case basis.

\subsection{Nature of the obscuration}

The origin of obscuration in AGN at early epochs is still unclear. In the local universe it has become widely accepted that this is done by a cold molecular torus surrounding the nucleus on parsec scales. However this picture is not so clear at higher redshifts \citep{draper11,page11}. Host galaxy and/or star forming processes have been suggested as possible obscuring scenarios, many of which stem from galaxy/AGN formation models \citep[e.g.][]{hopkins06}.

We have found evidence that the properties of the obscuring material is closely linked to the properties of the AGN. Figure \ref{fig_ftor} shows the fraction of spectra which show heavily obscured emission, but with no evidence for soft-scattered or Compton-reflected emission, indicative of a heavily buried source with a large covering factor. We find this to be a decreasing function of X-ray luminosity and interpret it as the average covering factor of the torus decreasing with \lx, consistent with the decline of the obscured fraction with \lx\ \citep[e.g.][]{ueda03} and the receding torus model \citep{lawrence91}. This strong relationship between the covering factor of the obscuring material and the luminosity of the AGN implies a close link between the AGN and its obscuration, in favour of a local origin. 

However, we also show that this `heavily buried' fraction increases with redshift. It is difficult to reconcile this observation with the properties of a torus, as this would imply an evolution in the torus covering factor which would be difficult to explain. Instead, we posit that this rising fraction of heavily buried sources may be due to obscuration on larger host-galaxy scales. Several recent works have found evidence for additional obscuration not linked to the AGN \citep[e.g.][]{juneau13, stern14}. Indeed this was found to be the case in the CT AGN XID403, where an ALMA observation found evidence for a compact nuclear starburst, the gas in which could plausibly be causing the measured CT obscuration \citep{gilli14}. This sample is ideally suited to similar follow-up studies to assess the connection between CT obscuration and the host galaxy, including further ALMA observations.

\section{Summary and Conclusions}
\label{sec_conc}

We have presented the results from the X-ray spectral analysis of \nred\ sources in the {\it Chandra} Deep Field-South, AEGIS-XD and {\it Chandra}-COSMOS surveys, with the primary goals of identifying Compton thick AGN and characterising obscuration across a wide range of redshifts and X-ray luminosities. For this we have used X-ray spectral torus models which self consistently account for the major signatures of absorption and thus give the best estimate of the line of sight obscuration and intrinsic X-ray luminosity for these heavily obscured sources. The results from this work can be found in online data tables at  \url{http://www.mpe.mpg.de/~mbright/data/} along with the X-ray spectral models used.

Our results are as follows:

\begin{itemize}
\item We present redshifts, \nh, $\Gamma$ (with uncertainties), observed X-ray fluxes and intrinsic 2-10 keV luminosities for all \nred\ sources analysed, which are made publicly available in an online catalogue.
\item  We find a total of \nctagn\ sources which have a best-fit \nh$>10^{24}$ \cmsq. However, due the low-count nature of many of these sources, the constraints on \nh\ are often not good, finding that between \ncctagn\ and \npctagn\ sources are consistent with \nh$=10^{24}$ \cmsq\ considering the upper and lower confidence limits on \nh.
\item We define a sample of sources that are `highly probable' CT AGN having a best-fit \nh$>10^{24}$ \cmsq\ and have \nh\ constrained above $10^{23.5}$ \cmsq. There are a total of \ngctagn\ of these CT AGN across the three fields, \ngctagna\ in CDFS, \ngctagnb\ in AEGIS and \ngctagnc\ in COSMOS, which range from \lx=$10^{42}\sim3\times10^{45}$ \ergs, from $z=0.1-4$.
\item Utilising X-ray torus models, we find evidence for a decline in the torus covering factor towards higher \lx, consistent with the receding torus model. 
\item From the same analysis, we also find that the covering factor increases with redshift for the same \lx, consistent with studies that have shown the increase in the obscured fraction towards higher redshifts. We suggest that this may be due to additional obscuration not carried out by the torus.
\end{itemize}

\section{Acknowledgements}

The authors would like to thank the anonymous referee for the careful reading and constructive criticism of our manuscript. We thank Johannes Buchner and Yoshihiro Ueda for useful discussions. We would like to thank the builders and operators of \chandra\ and all those involved in the CDFS, AEGIS and COSMOS surveys. We thank the Max Planck Society for the funds, work environment and resources to carry out this research. This research has made use of data obtained from the Chandra Data Archive and software provided by the Chandra X-ray Center (CXC) in the application packages CIAO.

\bibliographystyle{mn2e}
\bibliography{bibdesk}


\label{lastpage}
\end{document}